\documentclass[nofootinbib,preprint,preprintnumbers,a4paper,10pt]{revtex4}
\usepackage{multirow}
\usepackage{textcomp}
\usepackage{amsmath,graphicx,color,epsfig}

\usepackage{footnote}
\usepackage{ulem}
\usepackage{booktabs}
\usepackage{array}
\usepackage{amssymb}
\usepackage{subfigure}
\usepackage{longtable}
\usepackage{verbatim}
\usepackage{amsfonts}
\usepackage{hyperref}
\usepackage{cancel}

\begin{document}

\title{$CP$ asymmetries in the $\Lambda_c^+\to pK^0_S$ and $\Xi^+_c\to \Sigma^+K^0_S$ decays}

\author{Di Wang$^{1}$}\email{wangdi@hunnu.edu.cn}
\author{Si-Jia Wen$^{1}$}
\address{%
$^1$Department of Physics, Hunan Normal University, and Key Laboratory of Low-Dimensional Quantum Structures and Quantum Control of Ministry of Education, Changsha, 410081, China
}

\begin{abstract}
$CP$ asymmetry is a crucial element in interpreting the matter-antimatter asymmetry in the universe and searching for new physics beyond the Standard Model.
There are three types of $CP$ asymmetry in charmed hadron decays into neutral kaons: the $CP$ asymmetry in $K^0-\overline K^0$ mixing, the direct $CP$ asymmetry in charmed hadron decay, and the $CP$-violating effect induced by the interference between charmed hadron decay and mixing of final-state kaon mesons.
In this work, we study the $CP$ asymmetries in the $\Lambda_c^+\to pK^0_S$ and $\Xi^+_c\to \Sigma^+K^0_S$ decays.
The time-independent and time-integrated $\Gamma$-, $\alpha$-, $\beta$-, and $\gamma$-defined $CP$ asymmetries in the chain decay $\mathcal{B}_{c\overline 3}\to \mathcal{B}K(t)(\to \pi^{+}\pi^{-})$ are derived.
It is found that the $CP$ asymmetry in $K^0-\overline K^0$ mixing cancels out in the $\alpha$-, $\beta$-, and $\gamma$-defined $CP$ asymmetries.
The $U$-spin analysis shows that the amplitudes of the $\Lambda_c^+\to pK^0$, $\Lambda_c^+\to p\overline K^0$, $\Xi^+_c\to \Sigma^+ K^0$, and $\Xi^+_c\to \Sigma^+ \overline K^0$ modes are not independent.
The hadronic parameters determining $CP$ asymmetries in the $\Lambda_c^+\to pK^0_S$ and $\Xi^+_c\to \Sigma^+K^0_S$ decays could be extracted from the $K^0_S-K^0_L$ asymmetry and decay parameters $\alpha$, $\beta$, and $\gamma$ in these two decay modes.
We find the $CP$-violating effect induced by the interference between charmed hadron decay and neutral kaon mixing in the $\Xi^+_c\to \Sigma^+ K^0_S$ decay could reach to be $\mathcal{O}(10^{-3})$, which is several times larger than those in $D$ meson decays and at the same order as the $CP$ asymmetry in $K^0-\overline K^0$ mixing.
In contrast, the same term in the $\Lambda_c^+\to pK^0_S$ mode are one order of magnitude smaller.
Thus, the $\Xi^+_c\to \Sigma^+ K^0_S$ decay is a promising mode for observing $CP$ asymmetry in the charmed hadron sector and verifying the $CP$-violating effect induced by the interference between charm decay and neutral kaon mixing.

\end{abstract}

\maketitle
\section{Introduction}

$CP$ asymmetry is a significant subject in particle physics, as it is a crucial element for interpreting the matter-antimatter asymmetry in the universe \cite{Sakharov:1967dj} and provides a window for searching for new physics.
$CP$ asymmetry can be accommodated in the Standard Model (SM) by the Kobayashi-Maskawa (KM) mechanism \cite{Cabibbo:1963yz,Kobayashi:1973fv}.
$CP$ asymmetries have been well established in meson systems \cite{LHCb:2019hro,Belle:2001zzw,Christenson:1964fg,BaBar:2001ags}.
Very recently, the LHCb Collaboration reported the first observation of $CP$ violation in bottom baryon decays \cite{LHCb:2025ray}.
However, $CP$ asymmetry in charmed baryon decays has not yet been observed, although many  experimental efforts have been performed \cite{BESIII:2018ciw,LHCb:2017hwf,Belle:2021crz,Belle:2022uod}.
Compared to the bottom system, the charm system is suitable for detecting new physics in up-type quark decay.
The $CP$ violation in charmed baryon decays is suppressed by the Glashow-Iliopoulos-Maiani mechanism \cite{Glashow:1970gm}, making it easier for new physics to manifested.
Theoretically, predicting direct $CP$ asymmetries in charmed baryon decays is challenging due to large uncertainties in estimating penguin diagrams.
The QCD-inspired approaches do not work well in the charm scale.
The penguin topologies cannot be extracted from branching fractions because they are much smaller than the tree topologies.
At present, the theoretical approaches for calculating the direct $CP$ violation in charmed baryon decays rely on phenomenological model.
For instance, the direct $CP$ asymmetry in the charmed baryon sector is predicted to be of order $\mathcal{O}(10^{-4})$ in rescattering dynamics \cite{Jia:2024pyb,Wang:2025khg}.

$CP$ asymmetry also appears in the Cabibbo-favored (CF) and the doubly Cabibbo-suppressed (DCS) charmed hadron decays into neutral kaons.
The time-dependent and time-integrated $CP$ asymmetries in $D$ decays into the $K^0_S$ meson were studied in Ref.~\cite{Yu:2017oky}.
It was found that, in addition to the $CP$ violation in $K^0-\overline K^0$ mixing and the direct $CP$ violation in charm decay, there is a third $CP$-violating effect resulting from the interference between the DCS and CF amplitudes with the mixing of final-state neutral kaons.
Since there are no penguin contributions in the CF and DCS transitions, the perturbative parameters can be extracted from data and are easer to calculate in theory.
Charmed baryon decays involve different partial wave amplitudes, which could provide more complementary $CP$ observables than $D$ meson decays \cite{Wang:2022fih,Wang:2024qff}.
The $CP$ asymmetries in charmed baryon decays into neutral kaons were studied in Ref.~\cite{Wang:2017gxe}.
Due to limitation in experimental data, we did not well constrain the hadronic parameters that determine $CP$ asymmetries in previous work.
Thanks to recent measurements of the branching fraction of the $\Xi^+_c\to \Sigma^+K^0_S$ decay \cite{Belle-II:2025klu}, the decay parameter $\alpha$ and the $K^0_S-K^0_L$ asymmetry in the $\Lambda_c^+\to pK^0_{S,L}$ decays \cite{LHCb:2024tnq,BESIII:2024sfz}, we can analyze the hadronic parameters in the $\Lambda_c^+\to pK^0_S$ and $\Xi^+_c\to \Sigma^+K^0_S$ decays in the $U$-spin limit.

In this work, we derive the time-independent and time-integrated $\Gamma$-, $\alpha$-, $\beta$-, and $\gamma$-defined $CP$ asymmetries in the chain decay $\mathcal{B}_{c\overline 3}\to \mathcal{B}K(t)(\to \pi^{+}\pi^{-})$, and then extract the hadronic parameters that determine the $CP$ asymmetries in the $\Lambda_c^+\to pK^0_S$ and $\Xi^+_c\to \Sigma^+K^0_S$ decays based on $U$-spin analysis.
It is found that the $CP$-violating effect induced by the interference between charmed hadron decay and neutral kaon mixing in the $\Xi^+_c\to \Sigma^+ K^0_S$ decay could reach to be $\mathcal{O}(10^{-3})$, which is several times larger than that in $D$ meson decays and at the same order as the $CP$ asymmetry in $K^0-\overline K^0$ mixing.
However, the same term in the $\Lambda^+_c\to p K^0_S$ decay is much smaller due to constraints on the theoretical parameters.
Thus, the $\Xi^+_c\to \Sigma^+ K^0_S$ mode is a promising channel for observing $CP$ asymmetry in charmed hadron decays and verifying the $CP$-violating effect induced by the interference between charmed hadron decay and neutral kaon mixing.
To further constrain $CP$ asymmetries in the $\Lambda_c^+\to pK^0_S$ and $\Xi^+_c\to \Sigma^+ K^0_S$ modes, experimental measurements of the $K^0_S-K^0_L$ asymmetry and decay parameters $\alpha$, $\beta$, and $\gamma$ in these two channels are suggested.

The rest of this paper is organized as follows. In Sec.~\ref{cp}, we derive the time-dependent and time-integrated $\Gamma$- and $\alpha$-defined $CP$ asymmetries in charmed baryon decays into neutral kaons.
The phenomenological analysis of the $CP$ asymmetries in the $\Lambda_c^+\to pK^0_S$ and $\Xi^+_c\to \Sigma^+K^0_S$ modes is presented in Sec.~\ref{us}.
Sec.~\ref{co} is a brief summary.
The formulas of the time-dependent and time-integrated $\beta$- and $\gamma$-defined $CP$ asymmetries are presented in Appendix~\ref{beta}.

\section{$CP$ asymmetries in the $\mathcal{B}_{c\overline 3}\to \mathcal{B}K^0_S$ decays}\label{cp}
In this section, we analyze the $CP$ asymmetries in the $\mathcal{B}_{c\overline 3}\to \mathcal{B}K^0_S$ decays.
The mass eigenstates of neutral kaons, $K_S^0$ and $K_L^0$, are linear combinations of the flavor eigenstates,
 \begin{equation}\label{eq:KSKL}
|K_{S,L}^0\rangle  =   \frac{1+\epsilon}{\sqrt{2(1+|\epsilon|^2)}}|K^0\rangle\mp
\frac{1-\epsilon}{\sqrt{2(1+|\epsilon|^2)}}|\overline{K}^0\rangle,
 \end{equation}
where $\epsilon$ is a complex parameter characterizing the $CP$ asymmetry in kaon mixing with
$|\epsilon|=(2.228\pm0.011)\times10^{-3}$ and $\phi_{\epsilon}=43.52\pm0.05^{\circ}$ \cite{ParticleDataGroup:2024cfk}.
In experiments, a $K^0_S$ candidate is reconstructed from its decay into two charged pions at a time close to its lifetime.
Not only $K^0_S$, but also $K^0_L$ serve as intermediate states in the $\mathcal{B}_{c\overline 3}\to \mathcal{B}K^0_S$ decays through $K^0_S-K^0_L$ oscillation \cite{Grossman:2011zk}.
The $\mathcal{B}_{c\overline 3}\to \mathcal{B}\overline K^0$ and $\mathcal{B}_{c\overline 3}\to \mathcal{B}K^0$ decays are
Cabibbo-favored and doubly Cabibbo-suppressed transitions, respectively.
Their decay amplitudes can be written as
\begin{align}\label{eq:ampCFDCS}
\mathcal{S}(\mathcal{B}_{c\overline 3}\to \mathcal{B} \overline K^{0})&=\mathcal{T}_{CF}^{S} e^{i(\phi_{CF}+\delta_{CF}^{S})},\qquad \mathcal{P}(\mathcal{B}_{c\overline 3}\to \mathcal{B} \overline K^{0})=\mathcal{T}_{CF}^{P} e^{i(\phi_{CF}+\delta_{CF}^{P})},\nonumber\\
\mathcal{S}(\mathcal{B}_{c\overline 3}\to \mathcal{B}K^0)&=\mathcal{T}_{DCS}^{S}\, e^{i(\phi_{DCS}+\delta_{DCS}^{S})},\qquad \mathcal{P}(\mathcal{B}_{c\overline 3}\to \mathcal{B}K^0)=\mathcal{T}_{DCS}^{P}\, e^{i(\phi_{DCS}+\delta_{DCS}^{P})},
\end{align}
where $\mathcal{T}^{S,P}_{CF,\,DCS}$ are the magnitudes of the decay amplitudes, $\phi_{CF,\,DCS}$ are the weak phases, and $\delta^{S,P}_{CF,\,DCS}$ are the strong phases.
The superscripts $S$ and $P$ are used to distinguish the $S$- and $P$-wave amplitudes.
The amplitudes of the $\overline{\mathcal{B}}_{c\overline 3}\rightarrow \overline{\mathcal{B}} K^0$ and $\overline{\mathcal{B}}_{c\overline 3}\rightarrow \overline{\mathcal{B}} \,\overline K^0$ decays are
\begin{align}\label{eq:ampbar}
  \mathcal{S}(\overline{\mathcal{B}}_{c\overline 3}\rightarrow \overline{\mathcal{B}} K^0)  &= - \mathcal{T}^{S}_{CF}e^{i(-\phi_{CF}+\delta^{S}_{CF})},\qquad
\mathcal{P}(\overline{\mathcal{B}}_{c\overline 3}\rightarrow \overline{\mathcal{B}} K^0)  = - \mathcal{T}^{P}_{CF}e^{i(-\phi_{CF}+\delta^{P}_{CF})},\nonumber\\
  \mathcal{S}(\overline{\mathcal{B}}_{c\overline 3}\rightarrow \overline{\mathcal{B}} \,\overline K^0) &= - \mathcal{T}^{S}_{DCS}e^{i(-\phi_{DCS}+\delta^{S}_{DCS})},\qquad
\mathcal{P}(\overline{\mathcal{B}}_{c\overline 3}\rightarrow \overline{\mathcal{B}} \,\overline K^0) = - \mathcal{T}^{P}_{DCS}e^{i(-\phi_{DCS}+\delta^{P}_{DCS})}.
\end{align}
To express the $CP$ asymmetry formulas clearly, we write the ratios of the DCS and CF amplitudes as
\begin{equation}\label{r}
\mathcal{S}(\mathcal{B}_{c\overline 3}\to \mathcal{B} K^0)/\mathcal{S}(\mathcal{B}_{c\overline 3}\to \mathcal{B}\overline K^0) = r^S_{\mathcal{B}}\,e^{i(\phi+\delta^S_{\mathcal{B}})},\qquad \mathcal{P}(\mathcal{B}_{c\overline 3}\to \mathcal{B} K^0)/\mathcal{P}(\mathcal{B}_{c\overline 3}\to \mathcal{B}\overline K^0) = r^P_{\mathcal{B}}\,e^{i(\phi+\delta^P_{\mathcal{B}})}.
\end{equation}
where $r^{S,P}_{\mathcal{B}} = \mathcal{T}^{S,P}_{DCS}/\mathcal{T}^{S,P}_{CF}$, $\delta^{S,P}_{\mathcal{B}} = \delta^{S,P}_{DCS} - \delta^{S,P}_{CF}$, and
\begin{align}
\phi=\phi_{DCS}-\phi_{CF}=Arg\left[-{V_{cd}^{*}V_{us}/V_{cs}^{*}V_{ud}}\right]=(-6.2\pm 0.4)\times 10^{-4}.
\end{align}
The ratio of the Cabibbo-favored $P$- and $S$-wave amplitudes is
\begin{equation}\label{r2}
\mathcal{P}(\mathcal{B}_{c\overline 3}\to \mathcal{B}\overline K^0)/\mathcal{S}(\mathcal{B}_{c\overline 3}\to \mathcal{B}\overline K^0) = r_{\mathcal{B}}\,e^{i\delta_{\mathcal{B}}},
\end{equation}
where $r_{\mathcal{B}} = \mathcal{T}^{P}_{CF}/\mathcal{T}^{S}_{CF}$ and $\delta_{\mathcal{B}} = \delta^{P}_{CF} - \delta^{S}_{CF}$.

The time-dependent $CP$ asymmetry in the $\mathcal{B}_{c\overline 3}\to \mathcal{B}K^0_S$ mode is defined by
\begin{equation}\label{m1}
A_{CP}(t) \equiv\frac{\Gamma_{\pi\pi}(t)-\overline
\Gamma_{\pi\pi}(t)}{\Gamma_{\pi\pi}(t)+\overline\Gamma_{\pi\pi}(t)},
\end{equation}
where
\begin{align}
  \Gamma_{\pi\pi}(t)&\equiv\Gamma(\mathcal{B}_{c\overline 3}\to \mathcal{B}K(t)(\to \pi^{+}\pi^{-})), \nonumber\\
\overline\Gamma_{\pi\pi}(t)&\equiv\Gamma(\overline {\mathcal{B}}_{c\overline 3}\to \overline{\mathcal{B}}K(t)
(\to \pi^{+}\pi^{-})).
\end{align}
The decay width $\Gamma$ for the $\mathcal{B}_{c\overline 3}\to\mathcal{B}M$ decay can be written as
\begin{align}\label{br}
&\Gamma = \frac{p_c}{8\pi}\frac{(m_{\mathcal{B}_{c\overline 3}}+m_\mathcal{B})^2-m_M^2}
{m_{\mathcal{B}_{c\overline 3}}^2}(|\mathcal{S}|^2+|\mathcal{P}|^2),
\end{align}
where $p_c$ is the center of momentum (CM) in the rest frame of the initial baryon.
The time-dependent $CP$ asymmetry of the $\mathcal{B}_{c\overline 3}\to \mathcal{B}K(t)(\to \pi^+\pi^-)$ decay is derived as
\begin{align}\label{eq:ACPt}
 A_{CP}(t)\simeq
\big(A_{CP}^{\overline K^0}(t)+A_{CP}^{\rm dir}(t)+A_{CP}^{\rm int}(t)\big)/{D(t)},
\end{align}
with
\begin{align}\label{eq:AcpK0}
A_{CP}^{\overline K^0}(t)
&=2\,(1+r^2_{\mathcal{B}})\Big[e^{-\Gamma_{K^0_S}t}  \mathcal{R}e(\epsilon)-e^{-\Gamma_K t}
\Big(\mathcal{R}e(\epsilon)\cos(\Delta m_Kt)+\mathcal{I}m(\epsilon)\sin(\Delta m_Kt)\Big)\Big],
\end{align}
\begin{align}
\label{eq:Acpint}
A_{CP}^{\rm int}(t)
&=-4\,\big(r^S_{\mathcal{B}}\sin\delta^S_{\mathcal{B}}
+r^2_{\mathcal{B}}\,r^P_{\mathcal{B}}\sin\delta^P_{\mathcal{B}}\big)\cos\phi
\nonumber\\&~~~~~\times\Big[e^{-\Gamma_{K^0_S}t}\,
\mathcal{I}m(\epsilon)-e^{-\Gamma_K t}
\,\Big(\mathcal{I}m(\epsilon)\cos(\Delta m_Kt)-\mathcal{R}e(\epsilon)\sin(\Delta m_Kt)\Big)\Big],
\end{align}
\begin{align}\label{dir3}
A_{CP}^{\rm dir}(t)&=2\,e^{-\Gamma_{K^0_S}t}\,\big(r^S_{\mathcal{B}}\sin\delta^S_{\mathcal{B}}
+r^2_{\mathcal{B}}\,r^P_{\mathcal{B}}\sin\delta^P_{\mathcal{B}}\big)\sin\phi,
\end{align}
\begin{align}
D(t)&= e^{-\Gamma_{K^0_S}t}\big[1-2\,r^S_{\mathcal{B}}\cos\delta^S_{\mathcal{B}}\cos\phi
+r^2_{\mathcal{B}}\big(1-2\,r^P_{\mathcal{B}}\cos\delta^P_{\mathcal{B}}\cos\phi\big)\big],
\end{align}
where $\Gamma_K\equiv(\Gamma_{K^0_S}+\Gamma_{K^0_L})/2$,
and $\Delta m_K\equiv m_{K^0_L}-m_{K^0_S}$.
The first term in Eq.~\eqref{eq:ACPt}, which is independent of the hadronic parameters $r_{\mathcal{B}}$ and $\delta_{\mathcal{B}}$, represents the $CP$ asymmetry in neutral kaon mixing.
The second term represents the direct $CP$ asymmetry induced by interference between the tree-level CF and DCS amplitudes.
The third term represents the $CP$-violating effect induced by interference between the CF and DCS amplitudes with the neutral kaon mixing.

Measurements of $CP$ asymmetries depend on the time intervals selected in experiments.
The time-integrated $CP$ asymmetry can be derived from the time-dependent asymmetry by introducing a time-dependent function, $F(t)$, to account for relevant experimental effects,
\begin{equation}\label{z21}
 A_{CP}(t_1,t_2)=\frac{\int^{t_2}_{t_1}dt\,F(t)\,\Gamma_{\pi\pi}(t)-\int^{t_2}_{t_1}dtF(t)\,\,\overline
\Gamma_{\pi\pi}(t)}{\int^{t_2}_{t_1}dt\,F(t)\,\Gamma_{\pi\pi}(t)
+\int^{t_2}_{t_1}dt\,F(t)\,\overline\Gamma_{\pi\pi}(t)}= \frac{\int^{t_2}_{t_1}dt\,F(t)\,\big[A_{CP}^{\overline K^0}(t)+A_{CP}^{\text{dir}}(t)+A_{CP}^{\text{int}}(t)\big]}
 {\int^{t_2}_{t_1}dt\,F(t)\,D(t)}.
\end{equation}
In the above formula, we adopt the approximation from \cite{Grossman:2011zk}
\begin{equation}
  F(t)=\begin{cases}&1 ~~~~~ t_1<t<t_2,\\
  &0~~~~~t>t_2~~\text{or}~~t<t_1.
\end{cases}
\end{equation}
The time-integrated $CP$ asymmetry is reduced as
\begin{align}\label{int}
A_{CP}(t_1,t_2)&\simeq \frac{2\big(r^S_{\mathcal{B}}\sin\delta^S_{\mathcal{B}}
+r^2_{\mathcal{B}}\,r^P_{\mathcal{B}}\sin\delta^P_{\mathcal{B}}\big)\sin\phi}{1-2\,r^S_{\mathcal{B}}\cos\delta^S_{\mathcal{B}}\cos\phi
+r^2_{\mathcal{B}}\big(1-2\,r^P_{\mathcal{B}}\cos\delta^P_{\mathcal{B}}\cos\phi\big)} \nonumber\\&~+ \frac{2(1+r^2_{\mathcal{B}})\mathcal{R}e(\epsilon)-4\mathcal{I}m(\epsilon)\,
\big(r^S_{\mathcal{B}}\sin\delta^S_{\mathcal{B}}
+r^2_{\mathcal{B}}\,r^P_{\mathcal{B}}\sin\delta^P_{\mathcal{B}}\big)\cos\phi}{1-2\,r^S_{\mathcal{B}}\cos\delta^S_{\mathcal{B}}\cos\phi
+r^2_{\mathcal{B}}\big(1-2\,r^P_{\mathcal{B}}\cos\delta^P_{\mathcal{B}}\cos\phi\big)}\nonumber\\&~~~~ \times \Bigg[1- \frac{\big[c(t_1)-c(t_2)\big]+\frac{(1+r^2_{\mathcal{B}})\mathcal{I}m(\epsilon)+2\mathcal{R}e(\epsilon)\,
\big(r^S_{\mathcal{B}}\sin\delta^S_{\mathcal{B}}
+r^2_{\mathcal{B}}\,r^P_{\mathcal{B}}\sin\delta^P_{\mathcal{B}}\big)\cos\phi}{(1+r^2_{\mathcal{B}})\mathcal{R}e(\epsilon)-2\mathcal{I}m(\epsilon)\,
\big(r^S_{\mathcal{B}}\sin\delta^S_{\mathcal{B}}
+r^2_{\mathcal{B}}\,r^P_{\mathcal{B}}\sin\delta^P_{\mathcal{B}}\big)\cos\phi}
\big[s(t_1)-s(t_2)\big]}{\tau_S\Gamma \,(1+x^2)(e^{-t_1/\tau_S}-e^{-t_2/\tau_S})}\Bigg],
 \end{align}
in which $x\equiv\Delta m/\Gamma$, $c(t)=e^{- \Gamma t}[\cos(\Delta m t)-x\, \sin(\Delta m t)]$, and $s(t)=e^{- \Gamma t}[x \cos(\Delta m t)+ \sin(\Delta m t)]$.
The first term, which is independent of $t_{1,2}$, represents the direct $CP$ asymmetry in charm decays.
In the remaining part of Eq. \eqref{int}, the terms proportional to $r^{S,P}_{\mathcal{B}}$ represent the $CP$-violating effect $A^{\rm int}_{CP}(t_1,t_2)$, and those without $r^{S,P}_{\mathcal{B}}$ represent the $CP$ violation in neutral kaon mixing.
In the limitation of $t_1\ll \tau_S\ll t_2 \ll \tau_L$, we have $e^{-\Gamma t_1}=e^{-\Gamma_S t_1}=1$ and $e^{-\Gamma t_2}=e^{-\Gamma_S t_2}=0$.
Then the time-integrated $CP$ violation can be written as
\begin{align}\label{q4}
A_{CP}(t_1\ll \tau_S\ll t_2 \ll \tau_L)&\simeq \frac{2\big(r^S_{\mathcal{B}}\sin\delta^S_{\mathcal{B}}
+r^2_{\mathcal{B}}\,r^P_{\mathcal{B}}\sin\delta^P_{\mathcal{B}}\big)\sin\phi}{1-2\,r^S_{\mathcal{B}}\cos\delta^S_{\mathcal{B}}\cos\phi
+r^2_{\mathcal{B}}\big(1-2\,r^P_{\mathcal{B}}\cos\delta^P_{\mathcal{B}}\cos\phi\big)}\nonumber\\&~+ \frac{2(1+r^2_{\mathcal{B}})\mathcal{R}e(\epsilon)-4\mathcal{I}m(\epsilon)\,
\big(r^S_{\mathcal{B}}\sin\delta^S_{\mathcal{B}}
+r^2_{\mathcal{B}}\,r^P_{\mathcal{B}}\sin\delta^P_{\mathcal{B}}\big)\cos\phi}{1-2\,r^S_{\mathcal{B}}\cos\delta^S_{\mathcal{B}}\cos\phi
+r^2_{\mathcal{B}}\big(1-2\,r^P_{\mathcal{B}}\cos\delta^P_{\mathcal{B}}\cos\phi\big)}\nonumber\\&~~\times\Bigg[1-
\frac{2}{1+x^2}-\frac{(1+r^2_{\mathcal{B}})\mathcal{I}m(\epsilon)+2\mathcal{R}e(\epsilon)\,
\big(r^S_{\mathcal{B}}\sin\delta^S_{\mathcal{B}}
+r^2_{\mathcal{B}}\,r^P_{\mathcal{B}}\sin\delta^P_{\mathcal{B}}\big)\cos\phi}{(1+r^2_{\mathcal{B}})\mathcal{R}e(\epsilon)-2\mathcal{I}m(\epsilon)\,
\big(r^S_{\mathcal{B}}\sin\delta^S_{\mathcal{B}}
+r^2_{\mathcal{B}}\,r^P_{\mathcal{B}}\sin\delta^P_{\mathcal{B}}\big)\cos\phi}\frac{2x}{1+x^2}\Bigg].
 \end{align}
Under the approximations $\mathcal{R}e(\epsilon) / \mathcal{I}m(\epsilon) \simeq -y/x$ and $y\approx-1$ \cite{Grossman:2009mn}, we obtain
\begin{align}\label{x1}
 A_{CP}(t_1\ll \tau_S\ll t_2 \ll \tau_L)& \simeq \big(A_{CP}^{\overline K^0}+A_{CP}^{\text {dir}}+A_{CP}^{\text{int}}\big)/D,
\end{align}
where
\begin{align}
A_{CP}^{\overline K^0}=-2\mathcal{R}e(\epsilon)(1+r^2_{\mathcal{B}}),
\end{align}
\begin{align}
A_{CP}^{\text {dir}}=2\big(r^S_{\mathcal{B}}\sin\delta^S_{\mathcal{B}}
+r^2_{\mathcal{B}}\,r^P_{\mathcal{B}}\sin\delta^P_{\mathcal{B}}\big)\sin\phi,
\end{align}
\begin{align}
A_{CP}^{\text{int}}=-4\mathcal{I}m(\epsilon)\,
\big(r^S_{\mathcal{B}}\sin\delta^S_{\mathcal{B}}
+r^2_{\mathcal{B}}\,r^P_{\mathcal{B}}\sin\delta^P_{\mathcal{B}}\big)\cos\phi,
\end{align}
\begin{align}
D=1-2\,r^S_{\mathcal{B}}\cos\delta^S_{\mathcal{B}}\cos\phi
+r^2_{\mathcal{B}}\big(1-2\,r^P_{\mathcal{B}}\cos\delta^P_{\mathcal{B}}\cos\phi\big).
\end{align}
In general, the total $CP$ asymmetry is dominated by the $CP$ violation in neutral kaon mixing, $A_{CP}^{\overline K^0}\simeq -2\mathcal{R}e(\epsilon)\approx -3.23\times 10^{-3}$.
The direct $CP$ asymmetry, $A_{CP}^{\rm dir}$, and the $CP$-violating effect induced by the interference between charm decay and neutral kaon mixing, $A_{CP}^{\rm int}$, are natively predicted to be at the order of $10^{-5}$ and $10^{-4}$, respectively \cite{Yu:2017oky}.

We can also define the time-dependent $CP$ asymmetry using the decay parameter $\alpha$,
\begin{equation}\label{m11}
A^{\alpha}_{CP}(t) \equiv\frac{\alpha_{\pi\pi}(t)+\overline
\alpha_{\pi\pi}(t)}{2},
\end{equation}
where
\begin{align}
 \alpha(t)=\frac{|\mathcal{H}_{+\frac{1}{2}}(t)|^2
 -|\mathcal{H}_{-\frac{1}{2}}(t)|^2}
 {|\mathcal{H}_{+\frac{1}{2}}(t)|^2
 +|\mathcal{H}_{-\frac{1}{2}}(t)|^2},\qquad
 \overline\alpha(t)=\frac{|\overline{\mathcal{H}}_{+\frac{1}{2}}(t)|^2
 -|\overline{\mathcal{H}}_{-\frac{1}{2}}(t)|^2}
 {|\overline{\mathcal{H}}_{+\frac{1}{2}}(t)|^2
 +|\overline{\mathcal{H}}_{-\frac{1}{2}}(t)|^2},
\end{align}
with the helicity amplitudes defined as
\begin{align}
  \mathcal{H}_{\pm\frac{1}{2}}(t)&\equiv\frac{1}{\sqrt{2}}[\mathcal{S}(\mathcal{B}_{c\overline 3}\to \mathcal{B}K(t)(\to \pi^{+}\pi^{-}))\pm \mathcal{P}(\mathcal{B}_{c\overline 3}\to \mathcal{B}K(t)(\to \pi^{+}\pi^{-}))],\nonumber\\
  \overline{\mathcal{H}}_{\pm\frac{1}{2}}(t)&\equiv\frac{1}{\sqrt{2}}[\mathcal{S}(\overline{\mathcal{B}}_{c\overline 3}\to \overline{\mathcal{B}}K(t)(\to \pi^{+}\pi^{-}))\mp \mathcal{P}(\overline{\mathcal{B}}_{c\overline 3}\to \overline{\mathcal{B}}K(t)(\to \pi^{+}\pi^{-}))].
\end{align}
The time-dependent $\alpha$-defined $CP$ asymmetry in the $\mathcal{B}_{c\overline 3}\to \mathcal{B}K(t)(\to \pi^+\pi^-)$ decay is derived as
\begin{align}
 A_{CP}^\alpha(t)\simeq
\big(A_{CP}^{\alpha,\rm dir}(t)+A_{CP}^{\alpha,\rm int}(t)\big)/{D^\alpha(t)},
\end{align}
where
\begin{align}
 A_{CP}^{\alpha,\rm dir}(t) = -2\,e^{-\Gamma_{K^0_S} t}r_{\mathcal{B}}\big[r^S_\mathcal{B}\big(\sin(\delta_\mathcal{B}+\delta^S_\mathcal{B})
 +r^2_\mathcal{B}\sin(\delta_\mathcal{B}-\delta^S_\mathcal{B})\big)
 -r^P_\mathcal{B}\big(\sin(\delta_\mathcal{B}+\delta^P_\mathcal{B})+r^2_\mathcal{B}\sin(\delta_\mathcal{B}-\delta^P_\mathcal{B}) \big)\big]\sin\phi,
\end{align}
\begin{align}
A_{CP}^{\alpha,\rm int}(t)&=4\,e^{-\Gamma_{K^0_S} t}r_{\mathcal{B}}\Big[r^S_\mathcal{B} \big[-\mathcal{R}e(\epsilon) \cos(\delta_\mathcal{B}+\delta^S_\mathcal{B})+\mathcal{I}m(\epsilon) \sin(\delta_\mathcal{B}+\delta^S_\mathcal{B})
 \nonumber\\&\qquad+r^2_\mathcal{B}\big(\mathcal{R}e(\epsilon) \cos(\delta_\mathcal{B}-\delta^S_\mathcal{B})+\mathcal{I}m(\epsilon) \sin(\delta_\mathcal{B}-\delta^S_\mathcal{B})
 \big)\big]
 \nonumber\\&\qquad~~+r^P_\mathcal{B} \big[\mathcal{R}e(\epsilon) \cos(\delta_\mathcal{B}+\delta^P_\mathcal{B})-\mathcal{I}m(\epsilon) \sin(\delta_\mathcal{B}+\delta^P_\mathcal{B})
 \nonumber\\&\qquad~~~-r^2_\mathcal{B}\big(\mathcal{R}e(\epsilon) \cos(\delta_\mathcal{B}-\delta^P_\mathcal{B})+\mathcal{I}m(\epsilon) \sin(\delta_\mathcal{B}-\delta^P_\mathcal{B})
 \big)\big]
 \Big]
\nonumber\\&~~~~~+4\,e^{-\Gamma_K t}r_{\mathcal{B}}\Big[
r^S_\mathcal{B}\big[\mathcal{R}e(\epsilon)\big(\cos(\Delta m_K t)\cos(\delta_\mathcal{B}+\delta^S_\mathcal{B})+\sin(\Delta m_K t)\sin(\delta_\mathcal{B}+\delta^S_\mathcal{B})\big)\nonumber\\&\qquad
+\mathcal{I}m(\epsilon)\big(\sin(\Delta m_K t)\cos(\delta_\mathcal{B}+\delta^S_\mathcal{B})-\cos(\Delta m_K t)\sin(\delta_\mathcal{B}+\delta^S_\mathcal{B})\big)
\nonumber\\&\qquad~-r^2_\mathcal{B}\big(\mathcal{R}e(\epsilon)\big(\cos(\Delta m t)\cos(\delta_\mathcal{B}-\delta^S_\mathcal{B})-\sin(\Delta m_K t)\sin(\delta_\mathcal{B}-\delta^S_\mathcal{B})\big)\nonumber\\&\qquad
~~+\mathcal{I}m(\epsilon)\big(\sin(\Delta m_K t)\cos(\delta_\mathcal{B}-\delta^S_\mathcal{B})+\cos(\Delta m_K t)\sin(\delta_\mathcal{B}-\delta^S_\mathcal{B})\big)\big)\big]\nonumber\\&\qquad
~~~-r^P_\mathcal{B}\big[\mathcal{R}e(\epsilon)\big(\cos(\Delta m_K t)\cos(\delta_\mathcal{B}+\delta^P_\mathcal{B})+\sin(\Delta m_K t)\sin(\delta_\mathcal{B}+\delta^P_\mathcal{B})\big)\nonumber\\&\qquad
~~~~+\mathcal{I}m(\epsilon)\big(\sin(\Delta m_K t)\cos(\delta_\mathcal{B}+\delta^P_\mathcal{B})-\cos(\Delta m_K t)\sin(\delta_\mathcal{B}+\delta^P_\mathcal{B})\big)
\nonumber\\&\qquad~~~~~~-r^2_\mathcal{B}\big(\mathcal{R}e(\epsilon)\big(\cos(\Delta m_K t)\cos(\delta_\mathcal{B}-\delta^P_\mathcal{B})-\sin(\Delta m_K t)\sin(\delta_\mathcal{B}-\delta^P_\mathcal{B})\big)\nonumber\\&\qquad
~~~~~~~+\mathcal{I}m(\epsilon)\big(\sin(\Delta m_K t)\cos(\delta_\mathcal{B}-\delta^P_\mathcal{B})+\cos(\Delta m_K t)\sin(\delta_\mathcal{B}-\delta^P_\mathcal{B})\big)\big)\big]
\Big],
\end{align}
\begin{align}
D^\alpha(t)= e^{-\Gamma_{K^0_S} t}(1+r_\mathcal{B}^2)^2.
\end{align}
In the above formula, the first term represents the direct $CP$ violation, and the other terms represent the $CP$-violating effect induced by the interference between the neutral kaon mixing and charmed baryon decay.

The time-integrated $\alpha$-defined $CP$ asymmetry is
\begin{equation}\label{m12}
A^{\alpha}_{CP}(t_1,t_2) \equiv\frac{\alpha_{\pi\pi}(t_1,t_2)+\overline
\alpha_{\pi\pi}(t_1,t_2)}{2},
\end{equation}
where
\begin{align}
 \alpha(t_1,t_2)=\frac{\int_{t_1}^{t_2}dt\,|\mathcal{H}_{+\frac{1}{2}}(t)|^2
 -\int_{t_1}^{t_2}dt\,|\mathcal{H}_{-\frac{1}{2}}(t)|^2}
 {\int_{t_1}^{t_2}dt\,|\mathcal{H}_{+\frac{1}{2}}(t)|^2
 +\int_{t_1}^{t_2}dt\,|\mathcal{H}_{-\frac{1}{2}}(t)|^2},\qquad
 \overline\alpha(t_1,t_2)=\frac{\int_{t_1}^{t_2}dt\,|\overline{\mathcal{H}}_{+\frac{1}{2}}(t)|^2
 -\int_{t_1}^{t_2}dt\,|\overline{\mathcal{H}}_{-\frac{1}{2}}(t)|^2}
 {\int_{t_1}^{t_2}dt\,|\overline{\mathcal{H}}_{+\frac{1}{2}}(t)|^2
 +\int_{t_1}^{t_2}dt\,|\overline{\mathcal{H}}_{-\frac{1}{2}}(t)|^2}.
\end{align}
In the limitation of $t_1\ll \tau_S\ll t_2 \ll \tau_L$, the time-integrated $\alpha$-defined $CP$ violation can be written as
\begin{align}
 &A_{CP}^{\alpha}(t_1\ll \tau_S\ll t_2 \ll \tau_L) = \big(A^{\alpha,\rm dir}_{CP}+A^{\alpha,\rm int}_{CP}\big)/D^\alpha,
\end{align}
where
\begin{align}
A^{\alpha,\rm dir}_{CP} = -2\,r_{\mathcal{B}}\big[r^S_\mathcal{B}\big(\sin(\delta_\mathcal{B}+\delta^S_\mathcal{B})
 +r^2_\mathcal{B}\sin(\delta_\mathcal{B}-\delta^S_\mathcal{B})\big)
 -r^P_\mathcal{B}\big(\sin(\delta_\mathcal{B}+\delta^P_\mathcal{B})+r^2_\mathcal{B}\sin(\delta_\mathcal{B}-\delta^P_\mathcal{B}) \big)\big]\sin\phi,
\end{align}
\begin{align}
A^{\alpha,\rm int}_{CP} &= 4\,\mathcal{I}m(\epsilon)\,r_{\mathcal{B}}\big[r^S_\mathcal{B}\big(\sin(\delta_\mathcal{B}+\delta^S_\mathcal{B})
 +r^2_\mathcal{B}\sin(\delta_\mathcal{B}-\delta^S_\mathcal{B})\big)
 -r^P_\mathcal{B}\big(\sin(\delta_\mathcal{B}+\delta^P_\mathcal{B})+r^2_\mathcal{B}\sin(\delta_\mathcal{B}-\delta^P_\mathcal{B}) \big)\big]\nonumber\\&~~~
 +4\,\mathcal{R}e(\epsilon)\,r_{\mathcal{B}}\big[r^S_\mathcal{B}\big(\cos(\delta_\mathcal{B}+\delta^S_\mathcal{B})
 -r^2_\mathcal{B}\cos(\delta_\mathcal{B}-\delta^S_\mathcal{B})\big)
 -r^P_\mathcal{B}\big(\cos(\delta_\mathcal{B}+\delta^P_\mathcal{B})-r^2_\mathcal{B}\cos(\delta_\mathcal{B}-\delta^P_\mathcal{B}) \big)\big],
\end{align}
\begin{align}
D^\alpha=(1+r_\mathcal{B}^2)^2.
\end{align}
Note that $CP$ asymmetry in the $K^0-\overline K^0$ mixing cancels out in the $\alpha$-defined $CP$ asymmetry.
If the $\alpha$-defined $CP$ astmmetry is observed in experiments,
the $CP$ asymmetry involving charmed baryon decay will be confirmed.

We can also define the time-dependent and time-integrated  $CP$ asymmetries using the decay parameters $\beta$ and $\gamma$.
The formulas for the time-dependent and time-integrated $\beta$- and $\gamma$-defined $CP$ asymmetries are presented in Appendix~\ref{beta}.
One can find that the $CP$ asymmetry in $K^0-\overline K^0$ mixing also vanishes in the $\beta$- and $\gamma$-defined $CP$ asymmetries.
The $\Gamma$-, $\alpha$-, $\beta$-, and $\gamma$-defined $CP$ asymmetries are complementary.
We can avoid the suppression of $CP$ violation due to certain strong phases by measuring these complementary observables.

\section{Phenomenological analysis}\label{us}

In this section, we analyze the $CP$ asymmetries in the $\Lambda_c^+\to pK^0_S$ and $\Xi^+_c\to \Sigma^+K^0_S$ decays under $U$-spin symmetry.
The charmed baryons $\Lambda_c^+$ and $\Xi_c^+$ form a $U$-spin doublet with $(U,U_3) = (1/2,\pm 1/2)$.
The octets baryons $p$ and $\Sigma^+$ form another $U$-spin doublet with $(U,U_3) = (1/2,\pm 1/2)$.
The quantum numbers $(U,U_3)$ of the $K^0$ and $\overline K^0$ mesons are $(1,1)$ and $(1,-1)$, respectively.
The Hamiltonian for the Cabibbo-favored transition changes the $U$-spin and its third component as $(\Delta U,\Delta U_3)=(1,-1)$.
The Hamiltonian for the doubly Cabibbo-suppressed transition changes the $U$-spin and its third component as $(\Delta U,\Delta U_3)=(1,1)$.
The $U$-spin amplitudes for the $\Lambda_c^+\to pK^0$, $\Lambda_c^+\to p\overline K^0$, $\Xi_c^+\to \Sigma^+K^0$, and $\Xi_c^+\to \Sigma^+\overline K^0$ decays are derived as
\begin{align}\label{u}
  \mathcal{S(P)}(\Lambda_c^+\to pK^0) &=  \langle \frac{1}{2},\frac{1}{2};1,1 |1,1;\frac{1}{2},\frac{1}{2} \rangle \times V^*_{cd}V_{us} = \mathcal{S(P)}_{\frac{3}{2}}\times V^*_{cd}V_{us},\nonumber\\
  \mathcal{S(P)}(\Lambda_c^+\to p\overline K^0) &=  \langle \frac{1}{2},\frac{1}{2};1,-1 |1,-1;\frac{1}{2},\frac{1}{2} \rangle \times V^*_{cs}V_{ud} = \left(\frac{1}{3}\mathcal{S(P)}_{\frac{3}{2}}-\frac{2}{3}\mathcal{S(P)}_{\frac{1}{2}}\right)\times V^*_{cs}V_{ud},\nonumber\\
  \mathcal{S(P)}(\Xi_c^+\to \Sigma^+K^0) &=  \langle \frac{1}{2},-\frac{1}{2};1,1 |1,1;\frac{1}{2},-\frac{1}{2} \rangle \times V^*_{cd}V_{us} = \left(\frac{1}{3}\mathcal{S(P)}_{\frac{3}{2}}-\frac{2}{3}\mathcal{S(P)}_{\frac{1}{2}}\right)\times V^*_{cd}V_{us},\nonumber\\
  \mathcal{S(P)}(\Xi_c^+\to \Sigma^+\overline K^0) &=  \langle \frac{1}{2},-\frac{1}{2};1,-1 |1,-1;\frac{1}{2},-\frac{1}{2} \rangle \times V^*_{cs}V_{ud} =\mathcal{S(P)}_{\frac{3}{2}}\times V^*_{cs}V_{ud}.
\end{align}
If we define
\begin{align}\label{r3}
  r_{S}e^{i\delta_S} = \frac{\mathcal{S}_{\frac{3}{2}}}{\left(\frac{1}{3}
\mathcal{S}_{\frac{3}{2}}-\frac{2}{3}\mathcal{S}_{\frac{1}{2}}\right)},\qquad r_{P}e^{i\delta_P} = \frac{\mathcal{P}_{\frac{3}{2}}}{\left(\frac{1}{3}
\mathcal{P}_{\frac{3}{2}}-\frac{2}{3}\mathcal{P}_{\frac{1}{2}}\right)},
\end{align}
the hadronic parameters determining the $CP$ asymmetries, $r^{S,P}_{p}$, $r^{S,P}_{\Sigma^+}$, $\delta^{S,P}_{p}$, and $\delta^{S,P}_{\Sigma^+}$ can be written as
\begin{align}\label{xx}
r^{S,P}_{p}  = -\left|\frac{V^*_{cd}V_{us}}{V^*_{cs}V_{ud}}\right|
\times r_{S,P},\qquad
r^{S,P}_{\Sigma^+}&= -\left|\frac{V^*_{cd}V_{us}}{V^*_{cs}V_{ud}}\right|\times \frac{1}{r_{S,P}},\qquad \delta^{S,P}_{p} = -\delta^{S,P}_{\Sigma^+} = \delta_{S,P},
\end{align}
and $r_{\Sigma^+} = r_p\cdot r_P/r_S$, $\delta_{\Sigma^+} = \delta_{p}+\delta_P-\delta_S$ in the $U$-spin limit.

The $U$-spin relations for the $\Lambda_c^+\to pK^0$, $\Lambda_c^+\to p\overline K^0$, $\Xi_c^+\to \Sigma^+K^0$, and $\Xi_c^+\to \Sigma^+\overline K^0$ modes can also be derived using the topological diagram approach \cite{Wang:2024ztg} and $SU(3)$ irreducible amplitude approach \cite{Jia:2019zxi}, and by applying the operator $S=U_+-\lambda U_3-\lambda^2 U_-$ to the initial and final states of these decay channels \cite{Wang:2022kwe}.
The decay modes $\Lambda_c^+\to pK^0$ and $\Xi_c^+\to \Sigma^+\overline K^0$, as well as $\Lambda_c^+\to p\overline K^0$ and $\Xi_c^+\to \Sigma^+K^0$, are  $U$-spin conjugate channels.
The decay amplitudes of the $U$-spin conjugate CF and DCS channels are connected by the interchange of $V_{cd}^*V_{us}\leftrightarrow V_{cs}^*V_{ud}$.
This conclusion can be proven using the angular momentum coupling rule \cite{Varshalovich:1988ifq}:
\begin{align}\label{p}
 \langle j_1,-m_1;j_2,-m_2|j_{3},-m_{3};j_{4},-m_{4}\rangle=(-1)^{j_1+j_2-j_{3}-j_{4}}\langle &j_1,m_1;j_2,m_2|j_{3},m_{3};j_{4},m_{4}\rangle.
\end{align}
More detailed discussions about $U$-spin conjugate channels can be found in Refs.~\cite{Zhang:2025jnw,Wang:2019dls,Wen:2025ibn}.

The hadronic parameters $r_{S,P}$, $\delta_{S,P}$, $r_{p}$, and $\delta_{p}$ are constrained by the ratio between two branching fractions, the $K^0_S-K^0_L$ asymmetry, and the decay parameters in the $\Lambda_c^+\to pK^0_S$ and $\Xi_c^+\to \Sigma^+K^0_S$ decays.
The branching fractions $\mathcal{B}r(\Lambda_c^+\to pK^0_S)$ and $\mathcal{B}r(\Xi_c^+\to \Sigma^+K^0_S)$ are given by \cite{ParticleDataGroup:2024cfk,Belle-II:2025klu}
\begin{align}\label{d1}
  \mathcal{B}r(\Lambda_c^+\to pK^0_S) = (16.1\pm 0.7)\times 10^{-3},\qquad
  \mathcal{B}r(\Xi_c^+\to \Sigma^+K^0_S) = (1.94\pm 0.90)\times 10^{-3}.
\end{align}
According to Eqs.~\eqref{r2}, \eqref{br}, and \eqref{r3}, we have
\begin{align}
 \frac{|r_S|^2+|r_p|^2|r_P|^2}{1+|r_p|^2} =\frac{1+|r_{\Sigma^+}|^2}{|1/r_S|^2+|r_{\Sigma^+}|^2|1/r_P|^2} \simeq\frac{\kappa_{\Lambda^+_c}\tau_{\Lambda^+_c}}{\kappa_{\Xi^+_c}\tau_{\Xi^+_c}}
 \frac{\mathcal{B}r(\Xi_c^+\to \Sigma^+K^0_S)}{\mathcal{B}r(\Lambda_c^+\to pK^0_S)},
\end{align}
where
\begin{align}
 \kappa_{\Lambda^+_c} &=\frac{\sqrt{[m_{\Lambda^+_c}^2-(m_p+m_{K^0})^2][m_{\Lambda^+_c}^2-(m_p-m_{K^0})^2]}}{8\pi}\frac{(m_{\Lambda^+_c}+m_p)^2-m_{K^0}^2}
{2m_{\Lambda^+_c}^3},\\
 \kappa_{\Xi^+_c} &=\frac{\sqrt{[m_{\Xi^+_c}^2-(m_{\Sigma^+}+m_{K^0})^2][m_{\Xi^+_c}^2-(m_{\Sigma^+}-m_{K^0})^2]}}{8\pi}\frac{(m_{\Xi^+_c}+m_{\Sigma^+})^2-m_{K^0}^2}
{2m_{\Xi^+_c}^3}.
\end{align}
The $K_{S}^{0}-K_{L}^{0}$ asymmetry is defined as
\begin{equation}\label{Rf}
   R(\mathcal{B}_{c\overline 3}\to \mathcal{B}K^0_{S,L})\equiv\frac{\Gamma(\mathcal{B}_{c\overline 3}\to \mathcal{B}K^0_{S}) -\Gamma(\mathcal{B}_{c\overline 3}\to \mathcal{B}K^0_{L})}{\Gamma(\mathcal{B}_{c\overline 3}\to \mathcal{B}K^0_{S}) + \Gamma(\mathcal{B}_{c\overline 3}\to \mathcal{B}K^0_{L})},
 \end{equation}
 which can be expressed as
\begin{align}\label{Rfepsilon}
R(\mathcal{B}_{c\overline 3}\to \mathcal{B}K^0_{S,L}) \simeq -2\,\frac{r^S_\mathcal{B}\cos\delta^S_\mathcal{B}
+r^2_{\mathcal{B}}\,r^P_\mathcal{B}\cos\delta^P_\mathcal{B}}{1+r^2_{\mathcal{B}}}.
 \end{align}
The $K_{S}^{0}-K_{L}^{0}$ asymmetry in the $\Lambda^+_c\to pK^0_{S,L}$ decays is given by \cite{BESIII:2024sfz}
\begin{align}\label{d2}
  R(\Lambda^+_c\to pK^0_{S,L}) = (-2.5\pm 3.1)\times 10^{-2}.
\end{align}
Neglecting $CP$ violation, the decay parameters $\alpha$, $\beta$, and $\gamma$ in the $\mathcal{B}_{c\overline 3}\to \mathcal{B}K^0_{S}$ decay are approximately given by
\begin{align}
 \alpha(\mathcal{B}_{c\overline 3}\to \mathcal{B}K^0_{S}) \simeq \frac{2\,r_{\mathcal{B}}\cos\delta_{\mathcal{B}}}{1+r^2_{\mathcal{B}}},\qquad
 \beta(\mathcal{B}_{c\overline 3}\to \mathcal{B}K^0_{S}) \simeq \frac{2\,r_{\mathcal{B}}\sin\delta_{\mathcal{B}}}{1+r^2_{\mathcal{B}}},\qquad
 \gamma(\mathcal{B}_{c\overline 3}\to \mathcal{B}K^0_{S}) \simeq\frac{1-r^2_{\mathcal{B}}}{1+r^2_{\mathcal{B}}}.
\end{align}
Note that $\alpha^2+\beta^2+\gamma^2=1$.
The decay parameter $\alpha$ in the $\Lambda^+_c\to pK^0_{S}$ mode is given by \cite{LHCb:2024tnq}
\begin{align}\label{d3}
 \alpha(\Lambda^+_c\to pK^0_{S}) = -0.754\pm 0.010.
\end{align}
There are currently no data for the decay parameters $\beta$ and $\gamma$ in the $\Lambda^+_c\to pK^0_{S}$ and $\alpha$, $\beta$, $\gamma$ in the $\Xi_c^+\to \Sigma^+K^0_S$ modes.

\begin{figure}[!t]
\centering
\includegraphics[scale=0.37]{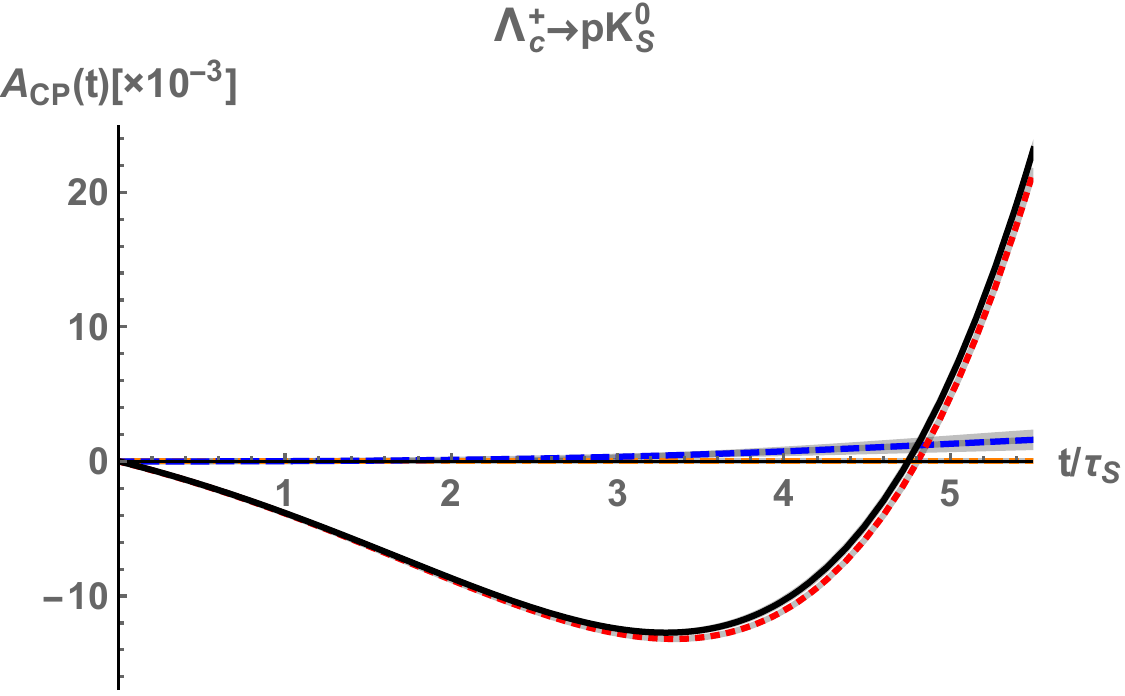}
\qquad\quad
\includegraphics[scale=0.37]{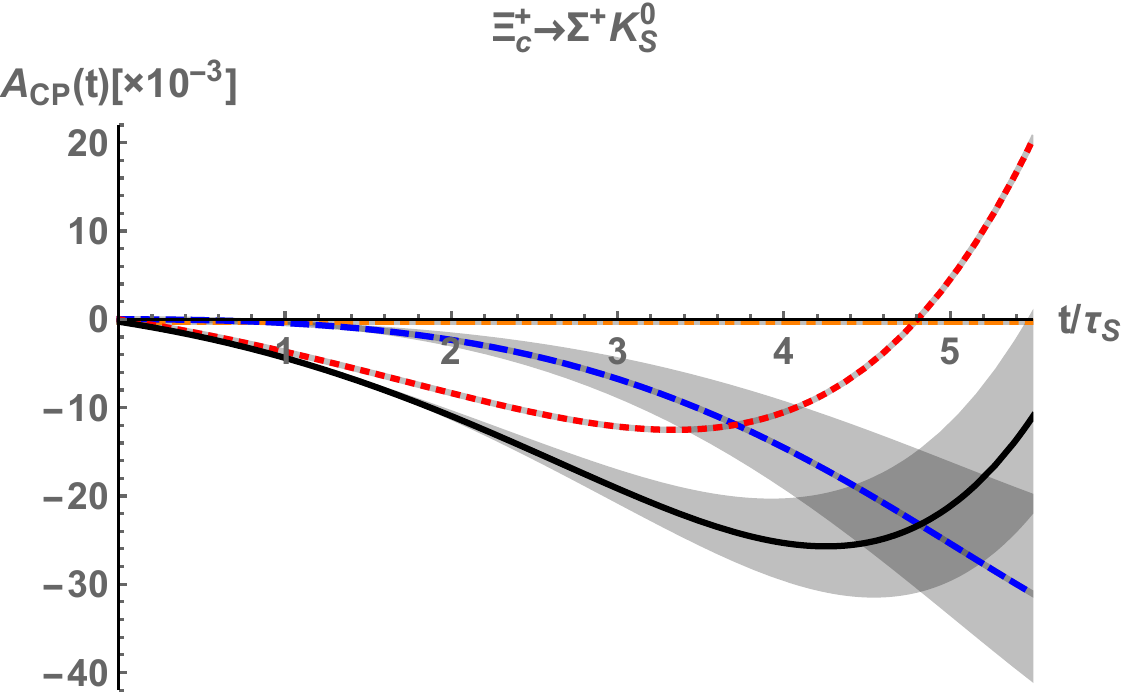}
\caption{Time-dependent $CP$ asymmetries in the $\Lambda^+_c\to p K(t)(\to \pi^+\pi^-)$ and $\Xi^+_c\to \Sigma^+ K(t)(\to \pi^+\pi^-)$ decays as functions of $t/\tau_S$.
The total $CP$ asymmetry is the black line, and the red, orange, and blue lines are the $A_{CP}^{\overline K^0}(t)$, $A_{CP}^{\rm int}(t)$, and $A_{CP}^{\rm dir}(t)$ terms, respectively.
The gray bands represent the theoretical uncertainties.
The ratio $r_p$, $r_{\Sigma^+}$ are set to $r_p =r_{\Sigma^+} = 1$, the relative strong phases $\delta^{S,P}_{p}$, $\delta^{S,P}_{\Sigma^+}$ are set to $\delta^{S,P}_{p} = -\delta^{S,P}_{\Sigma^+}=\pi/2$, and the ratios $r_p^{S,P}$, $r^{S,P}_{\Sigma^+}$ are taken from Eq.~\eqref{x}.
}
\label{cp1}
\end{figure}
\begin{figure}[!t]
\centering
\includegraphics[scale=0.37]{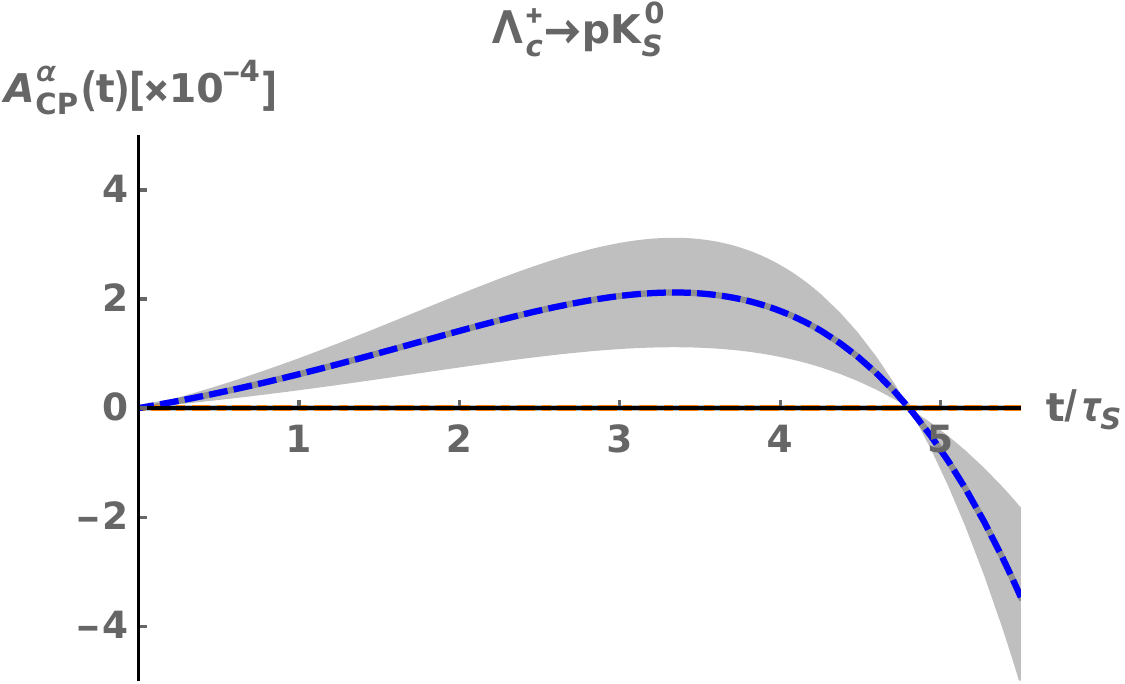}
\qquad\quad
\includegraphics[scale=0.37]{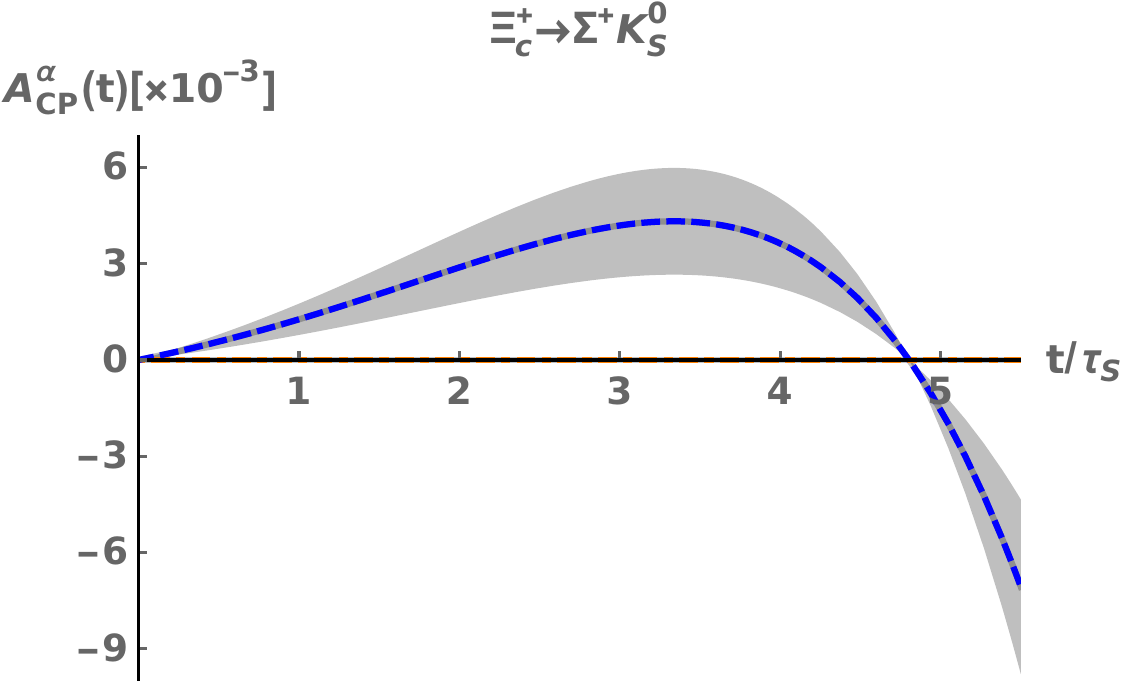}
\caption{Time-dependent $\alpha$-defined $CP$ asymmetries in the $\Lambda^+_c\to p K(t)(\to \pi^+\pi^-)$ and $\Xi^+_c\to \Sigma^+ K(t)(\to \pi^+\pi^-)$ decays as functions of $t/\tau_S$.
The orange and blue lines are the $A_{CP}^{\alpha,\rm int}(t)$ and $A_{CP}^{\alpha,\rm dir}(t)$ terms, respectively.
The ratio $r_p$, $r_{\Sigma^+}$ are set to $r_p =r_{\Sigma^+} = 1$, the relative strong phases $\delta^{S,P}_{p}$, $\delta^{S,P}_{\Sigma^+}$, $\delta_p$, and $\delta_{\Sigma^+}$ are set to $\delta^{S}_{p} = -\delta^{S}_{\Sigma^+}=\pi/2$, $\delta^{P}_{p} = -\delta^{P}_{\Sigma^+}=-\pi/2$, $\delta_p=3\pi/4$, and $\delta_{\Sigma^+}=-\pi/4$, and the ratios $r_p^{S,P}$, $r^{S,P}_{\Sigma^+}$ are taken from Eq.~\eqref{x}.
}
\label{cp2}
\end{figure}

Due to the limited experimental data, we cannot determine the theoretical parameters through global fitting.
In the special case where $ r_S= r_P$ and $r_p = 1$, the hadronic parameters $r^{S,P}_p$, $r^{S,P}_{\Sigma^+}$, $\delta_p$ are extracted to be
\begin{align}\label{x}
  r^{S,P}_p &= (-1.14\pm 0.26\pm 0.32\pm 0.34)\times 10^{-2},\qquad
  r^{S,P}_{\Sigma^+} = (-23.2\pm 5.4\pm 1.5\pm 7.0)\times 10^{-2},\nonumber\\
  \delta_p &= (0.77\pm 0.01\pm 0.02)\pi \quad{\rm or} \quad (1.23\pm0.01\pm 0.02)\pi.
\end{align}
In Eq.~\eqref{x}, the first uncertainty arises from experimental error.
The second uncertainty arises from neglecting the doubly Cabibbo-suppressed amplitudes in the $\Lambda_c^+\to pK^0_S$ and $\Xi_c^+\to \Sigma^+K^0_S$ decays.
The third uncertainty arises from $U$-spin breaking, which is naively expected to be $U_{\rm break}\sim m_s/\Lambda_{QCD}\sim 30\%$.
The ratios $r^{S,P}_p$ and $r^{S,P}_{\Sigma^+}$ in Eq.~\eqref{x} satisfy $r^{S,P}_p\times r^{S,P}_{\Sigma^+} \sim \lambda^4$, which is consistent with
Eq.~\eqref{xx}.
It is found in Eq.~\eqref{x} that $|r^{S,P}_p|\ll |r^{S,P}_{\Sigma^+}|$.
Actually, $|r^{S,P}_{\Sigma^+}|$ ($|r^{S,P}_{p}|$) is also larger (smaller) than the ratio between the DCS and CF amplitudes in $D$ meson decays into neutral kaons, $r_M$.
The ratios $r_{\pi^+}$ and $r_{K^+}$ in the $D^{+}\to \pi^{+}K_{S}^{0}$
and $D_{s}^{+}\to K^{+}K_{S}^{0}$ decays are estimated as \cite{Wang:2017ksn,Yu:2017oky}
\begin{equation}\label{eq:rdelta}
\begin{split}
   r_{\pi^+}=(-7.3\pm0.4)\times 10^{-2}, \qquad r_{K^+}=(-5.5\pm0.2)\times 10^{-2}.
\end{split}
\end{equation}
Thus, the direct $CP$ asymmetry and the $CP$-violating effect induced by the  interference between neutral kaon mixing and charmed hadron decay in the $\Xi_c^+\to \Sigma^+K^0_S$ mode could be larger than those in $D$ meson decays, while the same terms in the $\Lambda_c^+\to pK^0_S$ mode are smaller.
The time-dependent $\Gamma$-induced and $\alpha$-induced $CP$ asymmetries in the $\Lambda^+_c\to p K(t)(\to \pi^+\pi^-)$ and $\Xi^+_c\to \Sigma^+ K(t)(\to \pi^+\pi^-)$ decays, under the special case, as functions of $t/\tau_S$, are displayed in Figs.~\ref{cp1} and \ref{cp2}.
It is found that the $A_{CP}^{\rm int}(t)$ and $A_{CP}^{\alpha,\rm int}(t)$ terms in the $\Xi^+_c\to \Sigma^+ K(t)(\to \pi^+\pi^-)$ mode reach an order of $10^{-3}$ or even $10^{-2}$ in the range $ t \leq 5\tau_S$.
However, the same terms in the $\Lambda^+_c\to p K(t)(\to \pi^+\pi^-)$ mode are one order of magnitude smaller.

\begin{figure}[!t]
    \centering
\includegraphics[scale=0.35]{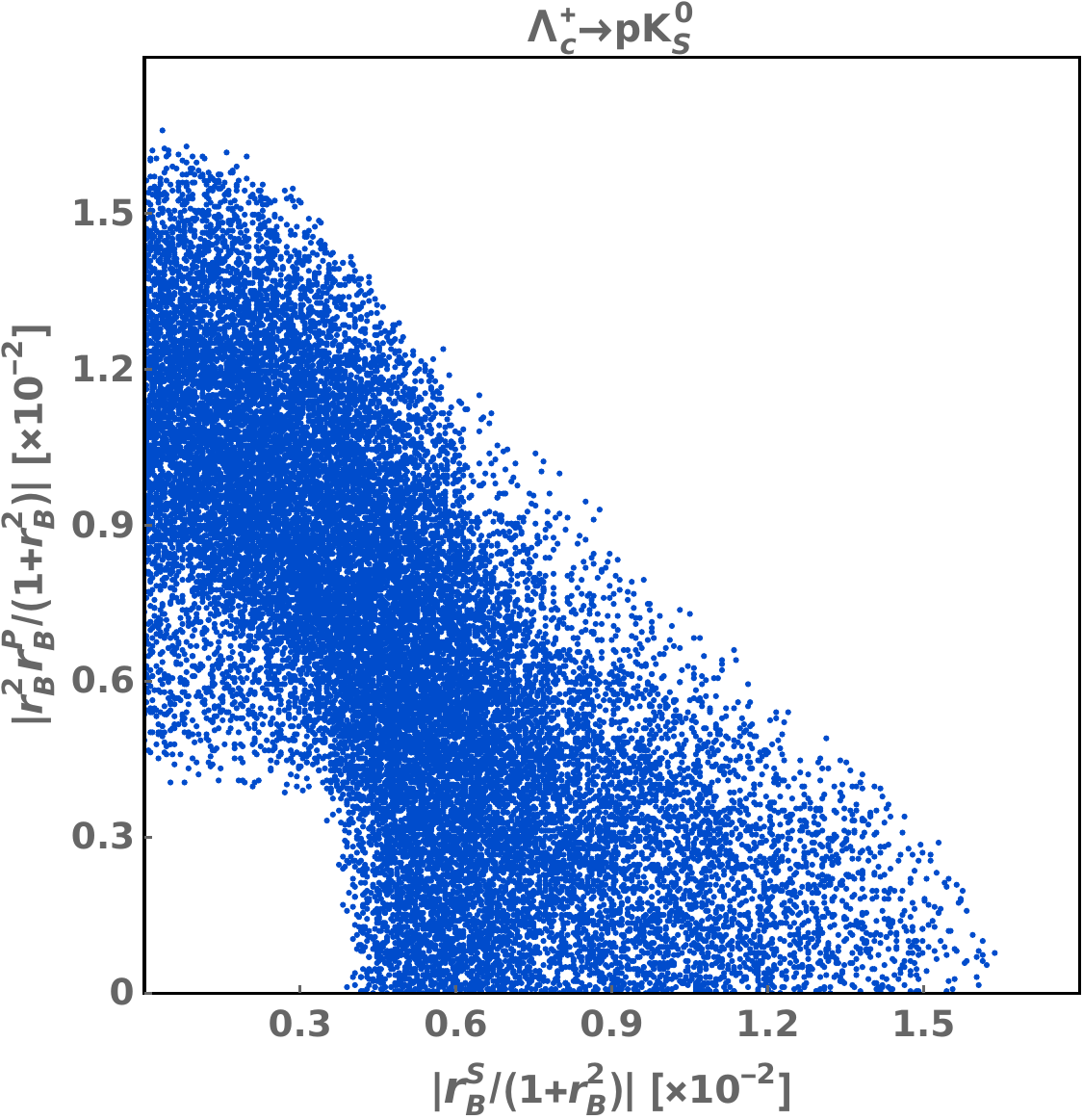}
\qquad\qquad
\includegraphics[scale=0.35]{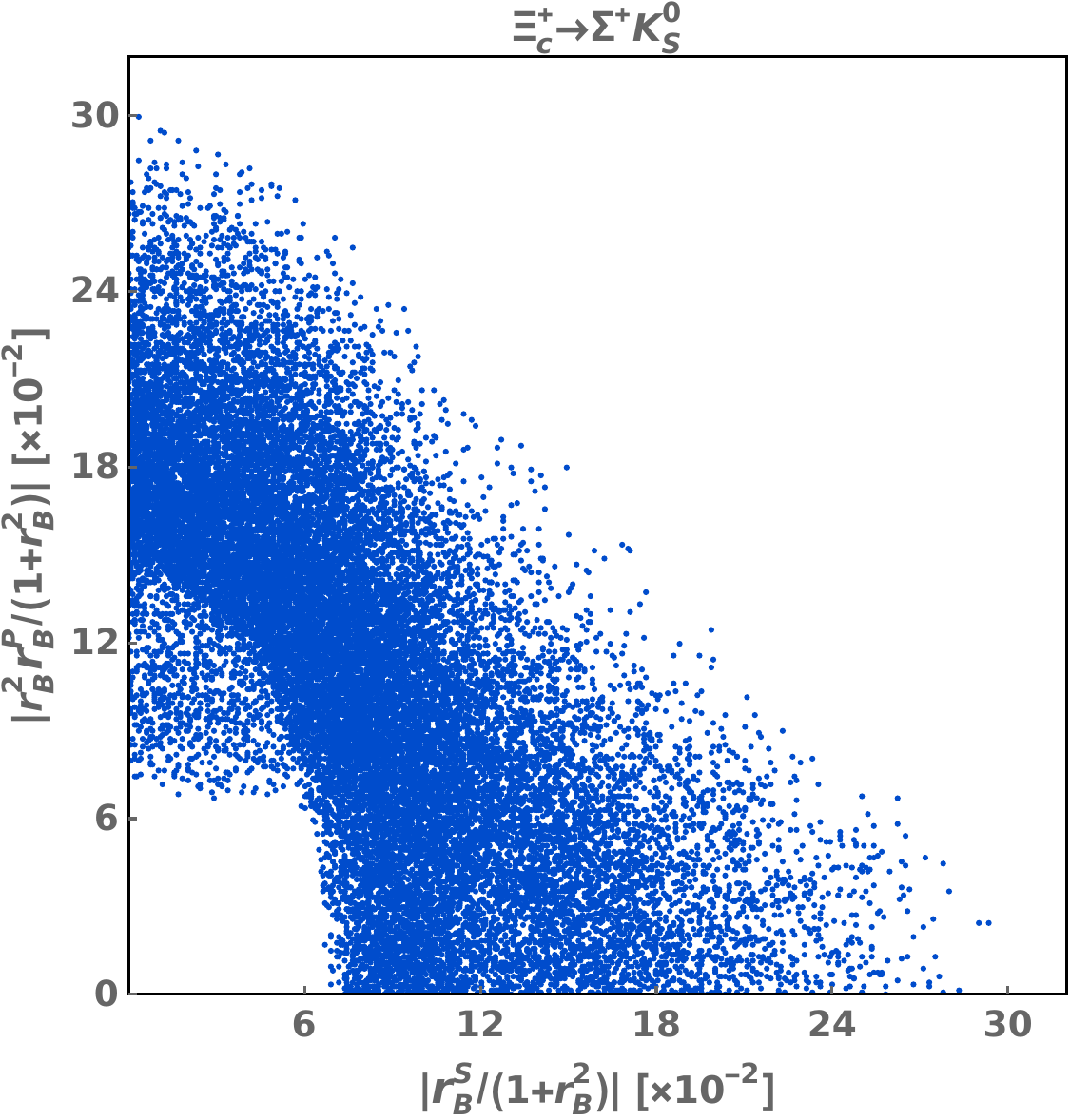}
\caption{The ranges of $r^S_\mathcal{B}/(1+r^2_\mathcal{B})$ and $r^2_\mathcal{B}\, r^P_\mathcal{B}/(1+r^2_\mathcal{B})$ in the $\Lambda^+_c\to p K^0_S$ and $\Xi^+_c\to \Sigma^+ K^0_S$ decays.
} \label{figr}
\end{figure}
\begin{figure}[!t]
    \centering
\includegraphics[scale=0.35]{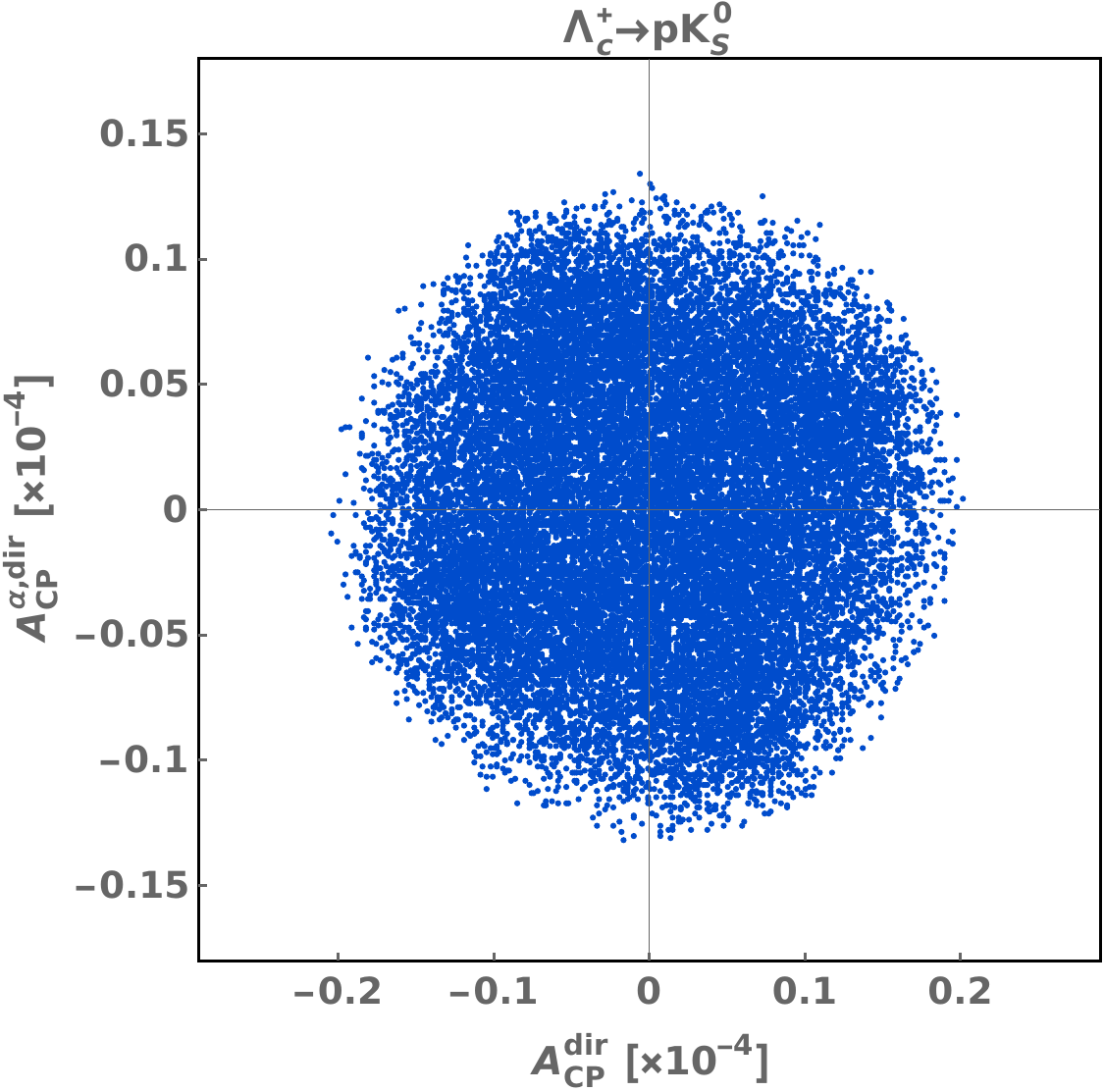}
\qquad\qquad
\includegraphics[scale=0.34]{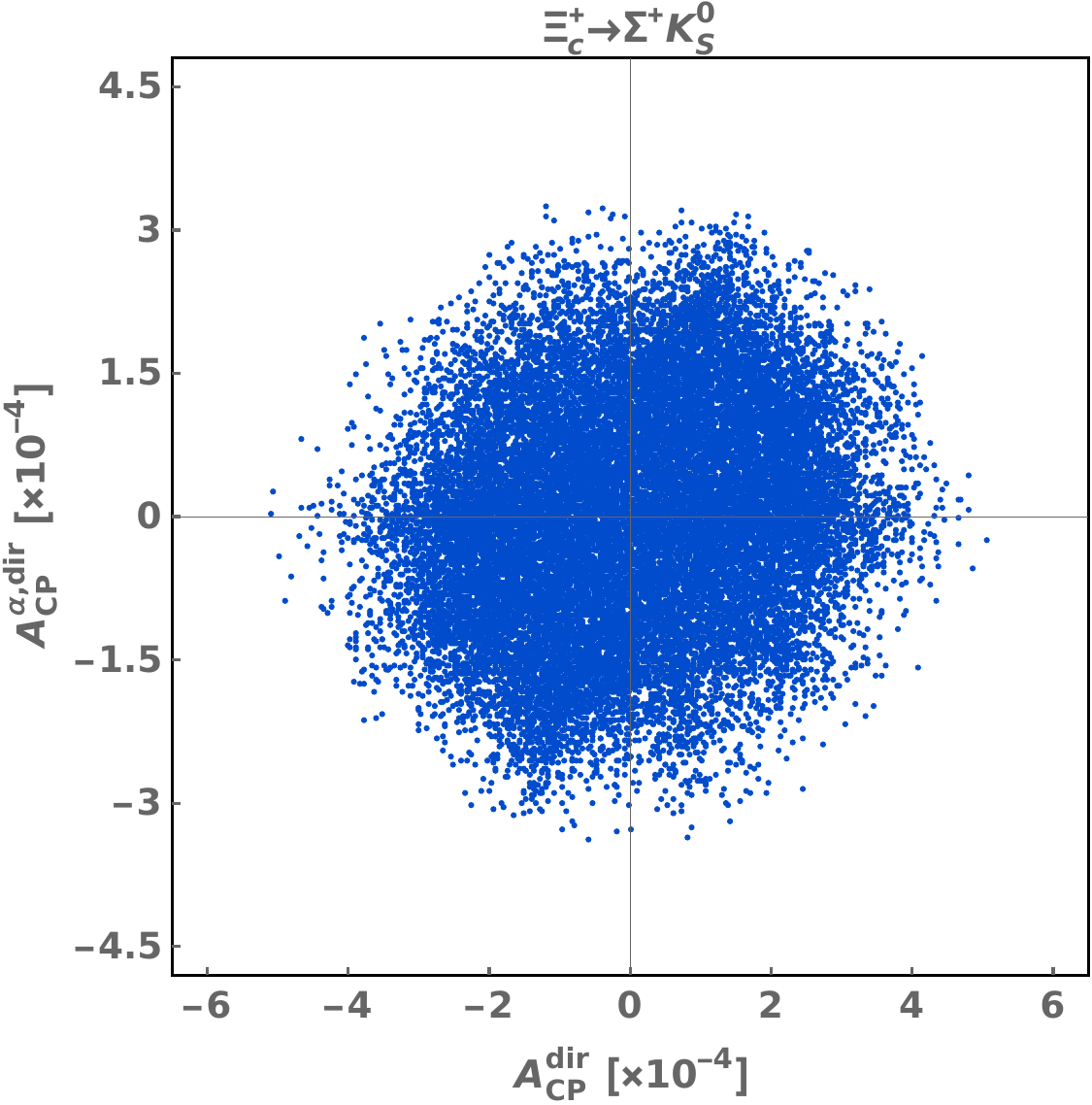}
\vspace{2mm}
\\
\includegraphics[scale=0.35]{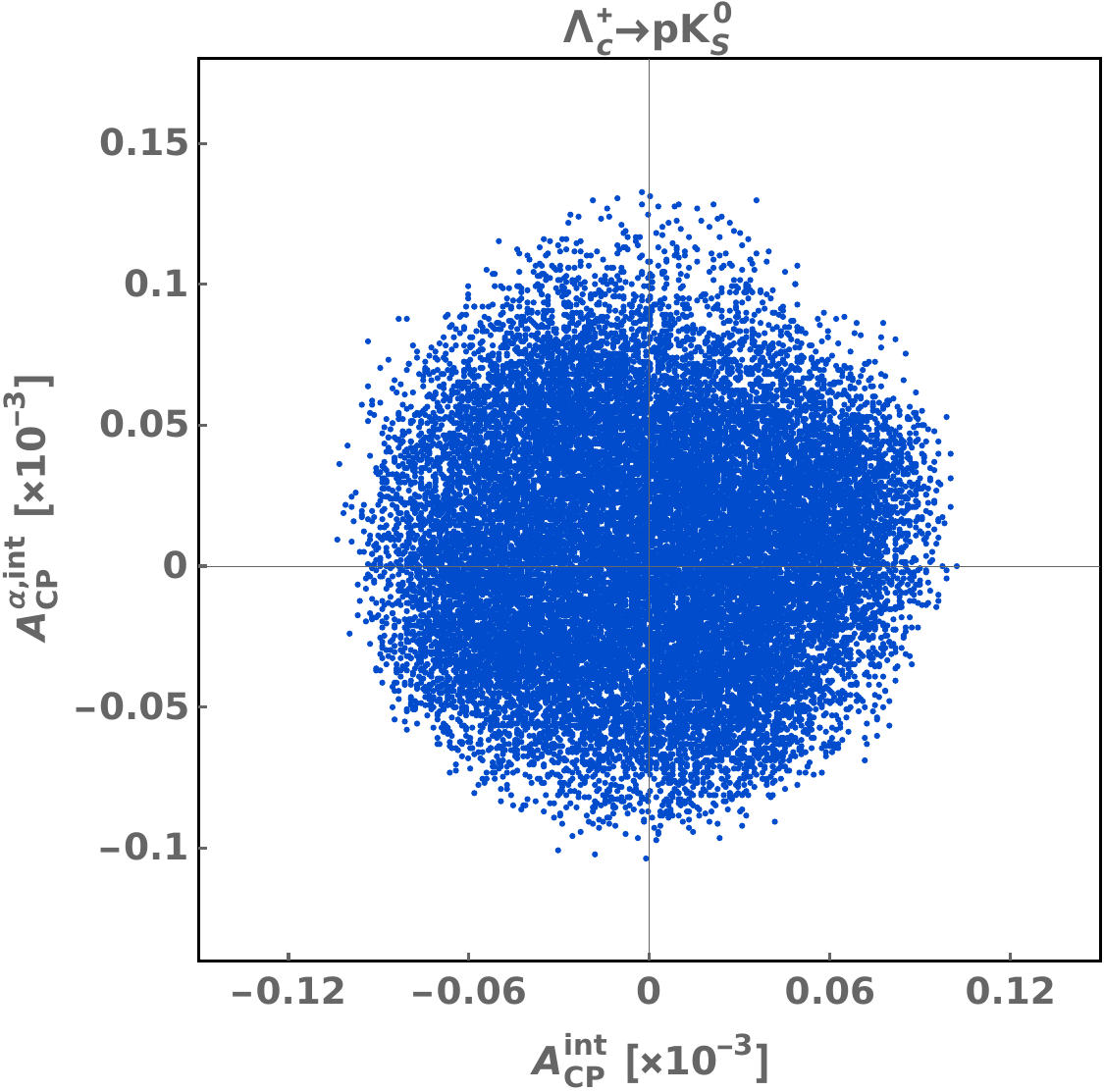}
\qquad\qquad
\includegraphics[scale=0.33]{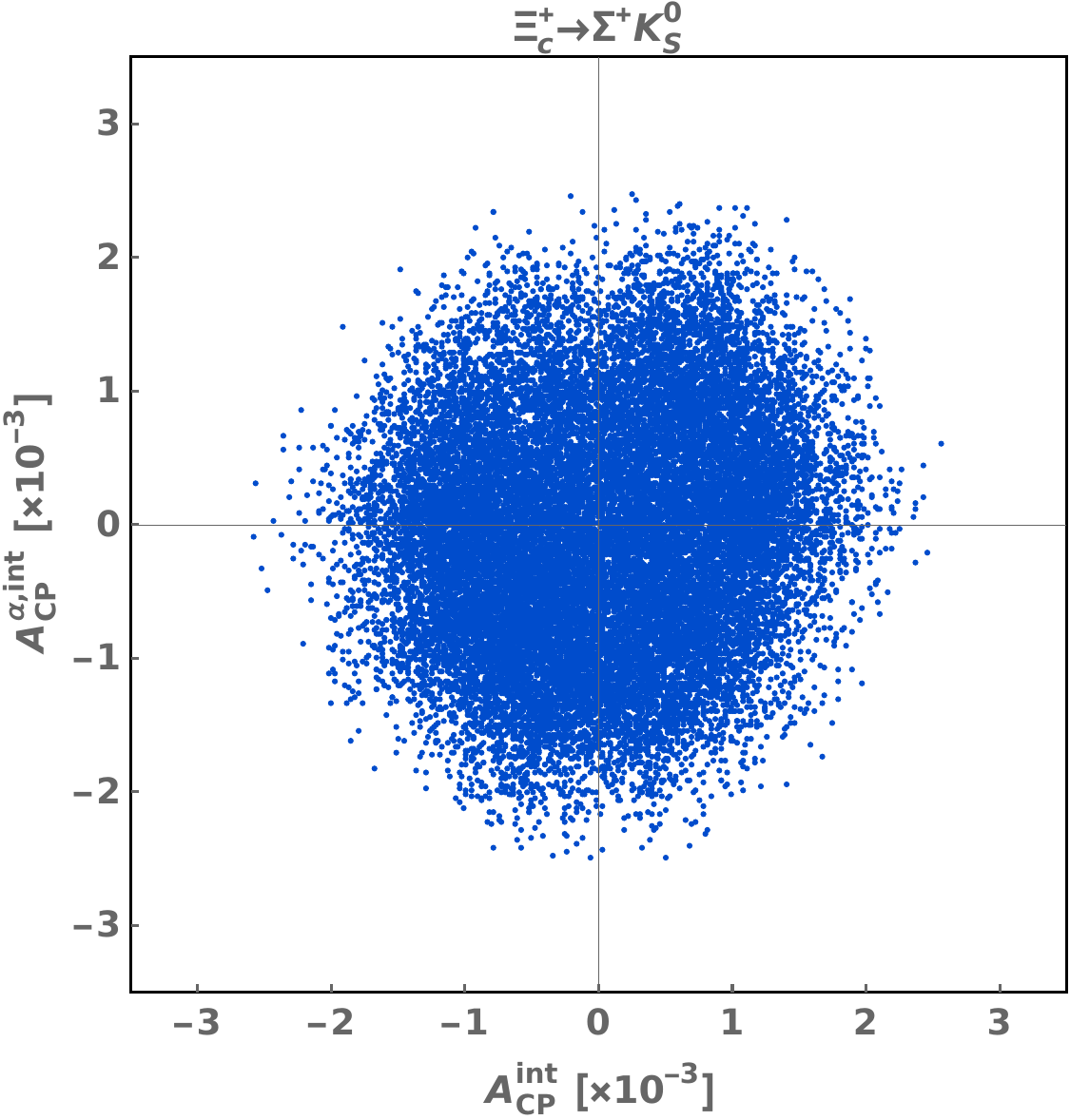}
\caption{The allowed ranges of time-integrated $\Gamma$- and $\alpha$-defined direct $CP$ asymmetries and $CP$-violating effects induced by the reference between $K^0-\overline K^0$ mixing and charmed baryon decay in the $\Lambda^+_c\to p K^0_S$ and $\Xi^+_c\to \Sigma^+ K^0_S$ decays.
} \label{cp4}
\end{figure}

With the experimental data of the observables $\mathcal{B}r(\Xi_c^+\to \Sigma^+K^0_S)/\mathcal{B}r(\Lambda_c^+\to pK^0_S)$, $R(\Lambda^+_c\to pK^0_{S,L})$, and $\alpha(\Lambda^+_c\to pK^0_{S})$,
we can constrain the allowed parameter space and limit the ranges of $CP$ asymmetries.
We define the $\chi^2$ as
\begin{align}
  \chi^2 = \sum_i\frac{({\rm Obs}^i_{\rm th}-{\rm Obs}^i_{\rm exp})^2}{(\Delta {\rm Obs}^i_{\rm exp})^2+(\Delta {\rm Obs}^i_{\rm th})^2},
\end{align}
where ${\rm Obs}^i_{\rm exp}$ and ${\rm Obs}^i_{\rm th}$ are the experimental and theoretical values of the observables, and $\Delta{\rm Obs}^i_{\rm exp}$ and $\Delta{\rm Obs}^i_{\rm th}$ are their experimental and theoretical uncertainties.
The theoretical uncertainty of $\mathcal{B}r(\Xi_c^+\to \Sigma^+K^0_S)/\mathcal{B}r(\Lambda_c^+\to pK^0_S)$ is dominated by $U$-spin breaking and the neglect of the DCS amplitudes, which is set to $50\%$ of the central value in this work.
The theoretical uncertainty of $\alpha(\Lambda^+_c\to pK^0_{S})$ is dominated by the neglect of the DCS amplitudes, which is estimated to be less than $0.03$.
The theoretical uncertainty of $R(\Lambda^+_c\to pK^0_{S,L})$ is dominated by the next to leading order contributions in the $r^S_p$ and $r^P_p$ expansions, which are negligible compared to the experimental uncertainty.
To constrain the parameter space, we sample the six theoretical parameters $r_{S,P}$, $r_{p}$, $\delta_{S,P}$, and $\delta_{p}$ within the ranges of $0\leq r_{S,P}\leq1$, $0\leq r_{p}\leq3$, $0\leq\delta_{S,P},\delta_{p}\leq2\pi$, retaining points where $\chi^2\leq1$.

The ranges of $r^S_\mathcal{B}/(1+r^2_\mathcal{B})$ and $r^2_\mathcal{B}\, r^P_\mathcal{B}/(1+r^2_\mathcal{B})$ in the $\Lambda^+_c\to p K^0_S$ and $\Xi^+_c\to \Sigma^+ K^0_S$ decays are shown in Fig.~\ref{figr}, as these two quantities determine the magnitudes of $CP$ asymmetries.
From Fig.~\ref{figr}, one can find that $r^S_\mathcal{B}/(1+r^2_\mathcal{B})$ and $r^2_\mathcal{B}\, r^P_\mathcal{B}/(1+r^2_\mathcal{B})$ in the $\Lambda^+_c\to p K^0_S$ mode are one order smaller than those in the $\Xi^+_c\to \Sigma^+ K^0_S$ in most areas.
The allowed ranges of time-integrated $\Gamma$-and $\alpha$-defined direct $CP$ violation and $CP$-violating effect induced by the interference between $K^0-\overline K^0$ mixing and charmed baryon decay in the $\Lambda^+_c\to p K^0_S$ and $\Xi^+_c\to \Sigma^+ K^0_S$ decays are shown in Fig.~\ref{cp4}.
Due to the small ratios $r^S_p/(1+r^2_p)$ and $r^2_p\, r^P_p/(1+r^2_p)$, the $A_{CP}^{\rm int}$ and $A_{CP}^{\alpha,\rm int}$ terms in the $\Lambda^+_c\to p K^0_S$ mode are restricted to an order of $10^{-4}$.
It is a challenging task to verifying $A_{CP}^{\rm int}$ and $A_{CP}^{\alpha,\rm int}$ in the $\Lambda^+_c\to p K^0_S$ mode in experiments so far.
Compared to the $\Lambda^+_c\to p K^0_S$ mode, the $A_{CP}^{\rm int}$ and $A_{CP}^{\alpha,\rm int}$ in the $\Xi^+_c\to \Sigma^+ K^0_S$ mode are at an order of $10^{-3}$ over most areas of the parameter space, which are much larger than those in charmed meson decay modes such as $D^+\to K^0_S\pi^+$ and $D^+_s\to K^0_SK^+$ \cite{Yu:2017oky}.
Thus, the $\Xi^+_c\to \Sigma^+ K^0_S$ decay is a promising mode to verify the $CP$ asymmetry induced by the interference between charmed hadron decay and neutral final-state kaon mixing.
Considering that the $CP$ asymmetries in most singly Cabibbo-suppressed charmed baryon decays are predicted to be $\mathcal{O}(10^{-4})$ \cite{Jia:2024pyb}, the $\Xi^+_c\to \Sigma^+ K^0_S$ decay is also a promising mode to observe $CP$ asymmetry in the charmed baryon sector in experiments.

The large ratios $r^S_{\Sigma^+}/(1+r^2_{\Sigma^+})$ and $r^2_{\Sigma^+} r^P_{\Sigma^+}/(1+r^2_{\Sigma^+})$ indicate a large $K^0_S-K^0_L$ asymmetry in the $\Xi^+_c\to \Sigma^+ K^0_S$ mode according to Eq.~\eqref{Rfepsilon}.
Measurements of the $K^0_S-K^0_L$ asymmetry $R(\Xi^+_c\to \Sigma^+ K^0_{S,L})$ could verify the doubly Cabibbo-suppressed $\Xi^+_c\to \Sigma^+ K^0$ mode.
Moreover, the $K^0_S-K^0_L$ asymmetries $R(\Lambda^+_c\to p K^0_{S,L})$ and $R(\Xi^+_c\to \Sigma^+ K^0_{S,L})$ are crucial observables for constraining the parameter space of $CP$ asymmetries in the $\Lambda^+_c\to p K^0_S$ and $\Xi^+_c\to \Sigma^+ K^0_{S}$ modes, as they could help us to extract the strong phases $\delta_S$ and $\delta_P$.
Besides, the decay parameters $\alpha$, $\beta$, and $\gamma$ in the $\Lambda^+_c\to pK^0_{S}$ and $\Xi^+_c\to \Sigma^+ K^0_S$ decays also provide restraints on the parameter space that determine $CP$ asymmetries.
If these observables are well measured by experiments, the $CP$ asymmetries in the $\Lambda^+_c\to pK^0_{S}$ and $\Xi^+_c\to \Sigma^+ K^0_S$ decays will be determined.

Eq.~\eqref{eq:ACPt} indicates that the total $CP$ asymmetry approaches  the direct $CP$ asymmetry at $t = 0$,
since both $A_{CP}^{\overline K^0}$ and $A_{CP}^{\rm int}$ are zero at $t = 0$.
Thus, the direct $CP$ asymmetry, $A_{CP}^{\rm dir}$, can be measured directly in experiments.
The direct $CP$ asymmetry could also be extracted by subtracting the $CP$ asymmetry in $K^0-\overline K^0$ mixing and the $CP$-violating effect induced by the interference between charmed hadron decay and neutral final-state kaon mixing from the total time-integrated $CP$ asymmetry.
If there is a large relative weak phase between the CF and DCS amplitudes induced by new physics, it could lead to an observable direct $CP$ asymmetry.
Such a $CP$-violating effect is more likely to be identified in the $\Lambda^+_c\to p K^0_S$ decay, as the direct $CP$ asymmetry in the SM is suppressed.
It is found in Fig.~\ref{cp4} that $|A_{CP}^{\rm dir}(\Lambda^+_c\to p K^0_S)|$ is smaller than $2.1\times 10^{-5}$ in the Standard Model.
Neither of the forthcoming experiments can attain such precision at an order of $10^{-5}$.
An observation of a nonvanishing $A_{CP}^{\rm dir}(\Lambda^+_c\to p K^0_S)$ would be a signature of new physics.
Compared to the $CP$ asymmetries in singly Cabibbo-suppressed decays and $D^0-\overline D^0$ mixing, the $\Lambda^+_c\to p K^0_S$ decay is not affected by loop diagrams, ensuring the reliability of the theoretical analysis.
Compared to other modes of charmed hadrons decaying into neutral kaons, the $CP$ asymmetry in the Standard Model and the uncertainties induced by non-perturbative QCD are suppressed by the small amplitude ratios in the $\Lambda^+_c\to p K^0_S$ mode,  enhancing the significance of new physics.
Thus, the $CP$ asymmetry in the $\Lambda^+_c\to p K^0_S$ decay is a potential window of searching for new physics in the charm sector.

\section{Summary}\label{co}

In summary, we studied $CP$ asymmetries in the $\Lambda_c^+\to pK^0_S$ and $\Xi^+_c\to \Sigma^+K^0_S$ decays.
The time-independent and time-integrated $CP$ asymmetries, defined using the decay width $\Gamma$ and the decay parameters $\alpha$, $\beta$, and $\gamma$ in charmed baryon decays into neutral kaons are derived.
The hadronic parameters that determine the $CP$ asymmetries in the $\Lambda_c^+\to pK^0_S$ and $\Xi^+_c\to \Sigma^+K^0_S$ decays are constrained by experimental data under $U$-spin symmetry.
It is found that the direct $CP$ asymmetry and the $CP$-violating effect induced by the interference between charmed hadron decay and neutral kaon mixing in the $\Xi_c^+ \to \Sigma^+ K_S^0$ decay could be several times larger than those in $D$ meson decays.
However, the same terms in the $\Lambda_c^+\to pK^0_S$ decay are one order of magnitude smaller.
Thus, the $\Xi^+_c\to \Sigma^+ K^0_S$ decay is a promising mode for observing $CP$ asymmetry in charmed baryon decays and verifying the $CP$-violating effect induced by the interference between charmed hadron decay and neutral kaon mixing.
To further constrain $CP$ asymmetries in the $\Lambda_c^+\to pK^0_S$ and $\Xi^+_c\to \Sigma^+ K^0_S$ modes, experimental measurements of the $K^0_S-K^0_L$ asymmetry and decay parameters $\alpha$, $\beta$, and $\gamma$ in these two channels are suggested.

\begin{acknowledgements}

We thank Pei-Rong Li for providing the motivation for this work and Jian-Peng Wang for suggesting modifying the definitions of the observables to avoid singularities.
This work was supported in part by the National Natural Science Foundation of China under Grants No. 12105099.

\end{acknowledgements}

\begin{appendix}
\section{$\beta$- and $\gamma$-defined $CP$ asymmetries}\label{beta}

The time-dependent $\beta$-defined $CP$ asymmetry is defined as
\begin{equation}
A^{\beta}_{CP}(t) \equiv\frac{\beta_{\pi\pi}(t)+\overline
\beta_{\pi\pi}(t)}{2},
\end{equation}
where
\begin{align}
 \beta(t)=\frac{2\,\mathcal{I}m[\mathcal{S}^*(t)
 \,\mathcal{P}(t)]}
 {|\mathcal{S}(t)|^2
 +|\mathcal{P}(t)|^2},\qquad
 \overline\beta(t)=-\frac{2\,\mathcal{I}m[\overline{\mathcal{S}}^*(t)
 \,\overline{\mathcal{P}}(t)]}
 {|\overline{\mathcal{S}}(t)|^2
 +|\overline{\mathcal{P}}(t)|^2}.
\end{align}
The time-dependent $\beta$-defined $CP$ asymmetry in the $\mathcal{B}_{c\overline 3}\to \mathcal{B}K(t)(\to \pi^+\pi^-)$ decay is derived as
\begin{align}
 A_{CP}^\beta(t)\simeq
\big(A_{CP}^{\beta,\rm dir}(t)+A_{CP}^{\beta,\rm int}(t)\big)/{D^\beta(t)},
\end{align}
where
\begin{align}
 A_{CP}^{\beta,\rm dir}(t) = 2\,e^{-\Gamma_{K^0_S} t}\,r_{\mathcal{B}}\big[r^S_\mathcal{B}\big(\cos(\delta_\mathcal{B}+\delta^S_\mathcal{B})
 +r^2_\mathcal{B}\cos(\delta_\mathcal{B}-\delta^S_\mathcal{B})\big)
 -r^P_\mathcal{B}\big(\cos(\delta_\mathcal{B}+\delta^P_\mathcal{B})
 +r^2_\mathcal{B}\cos(\delta_\mathcal{B}-\delta^P_\mathcal{B}) \big)\big]\sin\phi,
\end{align}
\begin{align}
A_{CP}^{\beta,\rm int}(t)&=-4\,e^{-\Gamma_{K^0_S} t}\,r_{\mathcal{B}}\Big[r^S_\mathcal{B} \big[\mathcal{R}e(\epsilon) \sin(\delta_\mathcal{B}+\delta^S_\mathcal{B})+\mathcal{I}m(\epsilon) \cos(\delta_\mathcal{B}+\delta^S_\mathcal{B})
 \nonumber\\&\qquad
 -r^2_\mathcal{B}\big(\mathcal{R}e(\epsilon) \sin(\delta_\mathcal{B}-\delta^S_\mathcal{B})-\mathcal{I}m(\epsilon) \cos(\delta_\mathcal{B}-\delta^S_\mathcal{B})
 \big)\big]
 \nonumber\\&\qquad~~
 -r^P_\mathcal{B} \big[\mathcal{R}e(\epsilon) \sin(\delta_\mathcal{B}+\delta^P_\mathcal{B})+\mathcal{I}m(\epsilon) \cos(\delta_\mathcal{B}+\delta^P_\mathcal{B})
 \nonumber\\&\qquad~~~
 +r^2_\mathcal{B}\big(\mathcal{R}e(\epsilon) \sin(\delta_\mathcal{B}-\delta^P_\mathcal{B})-\mathcal{I}m(\epsilon) \cos(\delta_\mathcal{B}-\delta^P_\mathcal{B})
 \big)\big]
 \Big]
\nonumber\\&~~~~~+4\,e^{-\Gamma_K t}\,r_{\mathcal{B}}\Big[
r^S_\mathcal{B}\big[-\mathcal{R}e(\epsilon)\big(\sin(\Delta m_K t)\cos(\delta_\mathcal{B}+\delta^S_\mathcal{B})-\cos(\Delta m_K t)\sin(\delta_\mathcal{B}+\delta^S_\mathcal{B})\big)
\nonumber\\&\qquad
+\mathcal{I}m(\epsilon)\big(\cos(\Delta m_K t)\cos(\delta_\mathcal{B}+\delta^S_\mathcal{B})+\sin(\Delta m_K t)\sin(\delta_\mathcal{B}+\delta^S_\mathcal{B})\big)
\nonumber\\&\qquad~
+r^2_\mathcal{B}\big(-\mathcal{R}e(\epsilon)\big(\sin(\Delta m_K t)\cos(\delta_\mathcal{B}-\delta^S_\mathcal{B})+\cos(\Delta m_K t)\sin(\delta_\mathcal{B}-\delta^S_\mathcal{B})\big)
\nonumber\\&\qquad
~~+\mathcal{I}m(\epsilon)\big(\cos(\Delta m_K t)\cos(\delta_\mathcal{B}-\delta^S_\mathcal{B})-\sin(\Delta m_K t)\sin(\delta_\mathcal{B}-\delta^S_\mathcal{B})\big)\big)\big]
\nonumber\\&\qquad
~~~+r^P_\mathcal{B}\big[\mathcal{R}e(\epsilon)\big(\sin(\Delta m_K t)\cos(\delta_\mathcal{B}+\delta^P_\mathcal{B})-\cos(\Delta m_K t)\sin(\delta_\mathcal{B}+\delta^P_\mathcal{B})\big)
\nonumber\\&\qquad
~~~~-\mathcal{I}m(\epsilon)\big(\cos(\Delta m_K t)\cos(\delta_\mathcal{B}+\delta^P_\mathcal{B})+\sin(\Delta m_K t)\sin(\delta_\mathcal{B}+\delta^P_\mathcal{B})\big)
\nonumber\\&\qquad~~~~~~
+r^2_\mathcal{B}\big(\mathcal{R}e(\epsilon)\big(\sin(\Delta m_K t)\cos(\delta_\mathcal{B}-\delta^P_\mathcal{B})+\cos(\Delta m_K t)\sin(\delta_\mathcal{B}-\delta^P_\mathcal{B})\big)
\nonumber\\&\qquad
~~~~~~~-\mathcal{I}m(\epsilon)\big(\cos(\Delta m_K t)\cos(\delta_\mathcal{B}-\delta^P_\mathcal{B})-\sin(\Delta m_K t)\sin(\delta_\mathcal{B}-\delta^P_\mathcal{B})\big)\big)\big]
\Big],
\end{align}
\begin{align}
D^\beta(t)= e^{-\Gamma_{K^0_S} t}(1+r_\mathcal{B}^2)^2.
\end{align}
In the above formula, the first term represents the direct $CP$ violation, and the other terms represent the $CP$ violation induced by the interference between the neutral kaon mixing and charmed baryon decay.
The time-integrated $\beta$-defined $CP$ asymmetry is
\begin{equation}
A^{\beta}_{CP}(t_1,t_2) \equiv\frac{\beta_{\pi\pi}(t_1,t_2)+\overline
\beta_{\pi\pi}(t_1,t_2)}{2},
\end{equation}
where
\begin{align}
\beta(t_1,t_2)=\frac{2\int_{t_1}^{t_2}dt\,\mathcal{I}m[\mathcal{S}^*(t)
 \,\mathcal{P}(t)]}
 {\int_{t_1}^{t_2}dt\,|\mathcal{S}(t)|^2
 +\int_{t_1}^{t_2}dt\,|\mathcal{P}(t)|^2},\qquad
 \overline\beta(t_1,t_2)=-\frac{2\int_{t_1}^{t_2}dt\,\mathcal{I}m[\overline{\mathcal{S}}^*(t)
 \,\overline{\mathcal{P}}(t)]}
 {\int_{t_1}^{t_2}dt\,|\overline{\mathcal{S}}(t)|^2
 +\int_{t_1}^{t_2}dt\,|\overline{\mathcal{P}}(t)|^2}.
\end{align}
In the limitation of $t_1\ll \tau_S\ll t_2 \ll \tau_L$, the time-integrated $\beta$-defined $CP$ violation can be written as
\begin{align}
 &A_{CP}^{\beta}(t_1\ll \tau_S\ll t_2 \ll \tau_L) = \big(A^{\beta,\rm dir}_{CP}+A^{\beta,\rm int}_{CP}\big)/D^\beta,
\end{align}
where
\begin{align}
A^{\beta,\rm dir}_{CP} = 2\,r_{\mathcal{B}}\big[r^S_\mathcal{B}\big(\cos(\delta_\mathcal{B}+\delta^S_\mathcal{B})
 +r^2_\mathcal{B}\cos(\delta_\mathcal{B}-\delta^S_\mathcal{B})\big)
 -r^P_\mathcal{B}\big(\cos(\delta_\mathcal{B}
 +\delta^P_\mathcal{B})+r^2_\mathcal{B}\cos(\delta_\mathcal{B}-\delta^P_\mathcal{B}) \big)\big]\sin\phi,
\end{align}
\begin{align}
A^{\beta,\rm int}_{CP} &= -4\,\mathcal{I}m(\epsilon)\,r_{\mathcal{B}}\big[r^S_\mathcal{B}\big(\cos(\delta_\mathcal{B}+\delta^S_\mathcal{B})
 +r^2_\mathcal{B}\cos(\delta_\mathcal{B}-\delta^S_\mathcal{B})\big)
 -r^P_\mathcal{B}\big(\cos(\delta_\mathcal{B}+\delta^P_\mathcal{B})
 +r^2_\mathcal{B}\cos(\delta_\mathcal{B}-\delta^P_\mathcal{B}) \big)\big]\nonumber\\&~~~
 +4\,\mathcal{R}e(\epsilon)\,r_{\mathcal{B}}\big[r^S_\mathcal{B}\big(\sin(\delta_\mathcal{B}+\delta^S_\mathcal{B})
 -r^2_\mathcal{B}\sin(\delta_\mathcal{B}-\delta^S_\mathcal{B})\big)
 -r^P_\mathcal{B}\big(\sin(\delta_\mathcal{B}+\delta^P_\mathcal{B})
 -r^2_\mathcal{B}\sin(\delta_\mathcal{B}-\delta^P_\mathcal{B}) \big)\big],
\end{align}
\begin{align}
D^\beta=(1+r_\mathcal{B}^2)^2.
\end{align}

The time-dependent $\gamma$-defined $CP$ asymmetry is defined as
\begin{equation}
A^{\gamma}_{CP}(t) \equiv\frac{\gamma_{\pi\pi}(t)-\overline
\gamma_{\pi\pi}(t)}{2},
\end{equation}
where
\begin{align}
 \gamma(t)=\frac{|\mathcal{S}(t)|^2
 -|\mathcal{P}(t)|^2}
 {|\mathcal{S}(t)|^2
 +|\mathcal{P}(t)|^2},\qquad
 \overline\gamma(t)=\frac{|\overline{\mathcal{S}}(t)|^2
 -|\overline{\mathcal{P}}(t)|^2}
 {|\overline{\mathcal{S}}(t)|^2
 +|\overline{\mathcal{P}}(t)|^2}.
\end{align}
The time-dependent $\gamma$-defined $CP$ asymmetry in the $\mathcal{B}_{c\overline 3}\to \mathcal{B}K(t)(\to \pi^+\pi^-)$ decay is derived as
\begin{align}
 A_{CP}^\gamma(t)\simeq
\big(A_{CP}^{\gamma,\rm dir}(t)+A_{CP}^{\gamma,\rm int}(t)\big)/{D^\gamma(t)},
\end{align}
where
\begin{align}
 A_{CP}^{\gamma,\rm dir}(t) = 4\,e^{-\Gamma_{K^0_S} t}r^2_\mathcal{B}\big(r^S_\mathcal{B}\sin\delta^S_\mathcal{B}
-r^P_\mathcal{B}\sin\delta^P_\mathcal{B}\big)\sin\phi,
\end{align}
\begin{align}
A_{CP}^{\gamma,\rm int}(t)&=8\,e^{-\Gamma_{K^0_S} t}r^2_{\mathcal{B}}\big[r^S_\mathcal{B} \big(\mathcal{R}e(\epsilon) \cos\delta^S_\mathcal{B}-\mathcal{I}m(\epsilon) \sin\delta^S_\mathcal{B}\big)
 -r^P_\mathcal{B} \big(\mathcal{R}e(\epsilon) \cos\delta^P_\mathcal{B}-\mathcal{I}m(\epsilon) \sin\delta^P_\mathcal{B}\big)
\big]
\nonumber\\
&~~-8\,e^{-\Gamma_K t}r^2_{\mathcal{B}}\Big[
r^S_\mathcal{B}\big[\mathcal{R}e(\epsilon)\big(\cos(\Delta m_K t)\cos\delta^S_\mathcal{B}+\sin(\Delta m_K t)\sin\delta^S_\mathcal{B}\big)
\nonumber\\&\quad+\mathcal{I}m(\epsilon)\big(\sin(\Delta m_K t)\cos\delta^S_\mathcal{B}-\cos(\Delta m_K t)\sin\delta^S_\mathcal{B}\big)\big]
\nonumber\\&\quad
~~~-r^P_\mathcal{B}\big[\mathcal{R}e(\epsilon)\big(\cos(\Delta m_K t)\cos\delta^P_\mathcal{B}+\sin(\Delta m_K t)\sin\delta^P_\mathcal{B}\big)
\nonumber\\&\quad\qquad+\mathcal{I}m(\epsilon)\big(\sin(\Delta m_K t)\cos\delta^P_\mathcal{B}-\cos(\Delta m_K t)\sin\delta^P_\mathcal{B}\big)\big]
\Big],
\end{align}
\begin{align}
D^\gamma(t)= e^{-\Gamma_{K^0_S} t}(1+r_\mathcal{B}^2)^2.
\end{align}
In the above formula, the first term represents direct $CP$ violation, and the other terms represent the $CP$ violation induced by the interference between the neutral kaon mixing and charmed baryon decay.
The time-integrated $\gamma$-defined $CP$ asymmetry is
\begin{equation}
A^{\gamma}_{CP}(t_1,t_2) \equiv\frac{\gamma_{\pi\pi}(t_1,t_2)-\overline
\gamma_{\pi\pi}(t_1,t_2)}{2},
\end{equation}
where
\begin{align}
\gamma(t_1,t_2)=\frac{\int_{t_1}^{t_2}dt\,|\mathcal{S}(t)|^2
 -\int_{t_1}^{t_2}dt\,|\mathcal{P}(t)|^2}
 {\int_{t_1}^{t_2}dt\,|\mathcal{S}(t)|^2
 +\int_{t_1}^{t_2}dt\,|\mathcal{P}(t)|^2},\qquad
 \overline\gamma(t_1,t_2)=\frac{\int_{t_1}^{t_2}dt\,|\overline{\mathcal{S}}(t)|^2
 -\int_{t_1}^{t_2}dt\,|\overline{\mathcal{P}}(t)|^2}
 {\int_{t_1}^{t_2}dt\,|\overline{\mathcal{S}}(t)|^2
 +\int_{t_1}^{t_2}dt\,|\overline{\mathcal{P}}(t)|^2}.
\end{align}
In the limitation of $t_1\ll \tau_S\ll t_2 \ll \tau_L$, the time-integrated $\gamma$-defined $CP$ violation can be written as
\begin{align}
 &A_{CP}^{\gamma}(t_1\ll \tau_S\ll t_2 \ll \tau_L) = \big(A^{\gamma,\rm dir}_{CP}+A^{\gamma,\rm int}_{CP}\big)/D^\gamma,
\end{align}
where
\begin{align}
A^{\gamma,\rm dir}_{CP} = 4\,r^2_\mathcal{B}\big(r^S_\mathcal{B}\sin\delta^S_\mathcal{B}
-r^P_\mathcal{B}\sin\delta^P_\mathcal{B}\big)\sin\phi,
\end{align}
\begin{align}
A^{\gamma,\rm int}_{CP} &= -8\,r^2_\mathcal{B}\big[\mathcal{I}m(\epsilon)\big(r^S_\mathcal{B}\sin\delta^S_\mathcal{B}
-r^P_\mathcal{B}\sin\delta^P_\mathcal{B}\big)
+\mathcal{R}e(\epsilon)\big(r^S_\mathcal{B}\cos\delta^S_\mathcal{B}
-r^P_\mathcal{B}\cos\delta^P_\mathcal{B}\big)\big],
\end{align}
\begin{align}
D^\gamma=(1+r_\mathcal{B}^2)^2.
\end{align}

\end{appendix}


\end{document}